%% file: aa54594-25.tex
\documentclass[longauth]{aaEC}
\makeatletter
\renewcommand*\aa@pageof{, page \thepage{} of \pageref*{LastPage}}
\makeatother
\usepackage{lastpage}
\usepackage{graphicx}
\usepackage{scalerel}
\usepackage[table]{xcolor}

\definecolor{lightgray}{gray}{0.95}
\usepackage{listings}
\usepackage{xcolor}

\usepackage[multiple]{footmisc}
\usepackage{xurl} 

\usepackage{txfonts}
\usepackage[pdfencoding=auto,psdextra]{hyperref}
\hypersetup{
    colorlinks=true,
    breaklinks=true, 
    linkcolor=blue,
    filecolor=magenta,      
    urlcolor=blue,
    citecolor=blue
}
\newcommand{\fnurl}[2]{#1\footnote{\href{#2}{\url{#2}}}}

\usepackage[utf8]{inputenc}

\usepackage[switch, modulo]{lineno}

\usepackage{euclid}
\usepackage[nameinlink,capitalise]{cleveref}
\crefname{section}{Sect.}{Sects.}
\Crefname{section}{Section}{Sections}
\crefname{figure}{Fig.}{Figs.}
\Crefname{figure}{Figure}{Figures}
\crefname{equation}{Eq.}{Eqs.}
\Crefname{equation}{Equation}{Equations}
\crefname{table}{Table}{Tables}
\crefname{appendix}{Appendix}{Appendices}

\usepackage[nolist,nohyperlinks]{acronym}

\begin{document}

\input{acronym}

\title{\Euclid Quick Data Release (Q1)}
\subtitle{VIS processing and data products}

\maxdeadcycles=500 

\input{authors}

\date\today 

\abstract{This paper describes the \ac{VIS PF} of the \Euclid ground segment pipeline, which processes and calibrates raw data from the VIS camera. We present the algorithms used in each processing element along with a description of the on-orbit performance of \ac{VIS PF} based on performance verification and Q1 datasets. We demonstrate that the principal performance metrics (image quality, astrometric accuracy, photometric calibration) are within pre-launch specifications. The image-to-image photometric scatter is less than \SI{0.8}{\percent} and absolute astrometric accuracy compared to Gaia is 5\,mas. Image quality is stable over all Q1 images, with a \ac{FWHM} of \ang{;;0.16}. The stacked images (combining four nominal and two short exposures) reach $\IE=25.6$ ($10\,\sigma$, measured as the variance of \ang{;;1.3} diameter apertures). We also describe quality control metrics provided with each image, and an appendix provides a detailed description of the provided data products. The excellent quality of these images demonstrates the immense potential of \Euclid VIS data for weak lensing. VIS data covering most of the extragalactic sky will provide a lasting high-resolution atlas of the Universe.  
}


   \titlerunning{Q1: VIS processing and data products}
   \authorrunning{Euclid collaboration: H.~J.~McCracken et al.}
   
   \maketitle

\section{\label{sc:Intro}Introduction}

The ESA \Euclid mission \citep{Laureijs11}, launched in July 2023, aims to significantly improve constraints of the dark energy equation of state \citep{EuclidSkyOverview}. \Euclid will achieve this by combining near-infrared spectroscopic redshifts that trace structures on large scales across cosmic time with optical galaxy-shape measurements that provide information on the distribution of dark matter. 

At the heart of \Euclid are two instruments, VIS \citep{EuclidSkyVIS} and NISP \citep{EuclidSkyNISP}, which can observe the sky simultaneously in the optical and the near-infrared using  a dichroic beam-splitter. The \ac{EWS} described in \citealt{Scaramella-EP1} uses these instruments to cover the darkest $14\,000\,\mathrm{deg}^2$ of the extragalactic sky, producing tens of thousands of images. The shapes and positions of billions of galaxies are extracted from these images using automated pipelines that run in a distributed processing environment known as the `Science Ground Segment'. The \ac{EDS} and auxiliary calibration fields comprise a set of smaller sky areas that are observed to forty times the depth of the \ac{EWS}. These fields provide an essential check for the quality of the ground segment processing. 

\Euclid's ambitious requirements place stringent constraints on the number of galaxies that must be observed, the accuracy with which their shapes can be measured, and therefore our knowledge of the telescope's \ac{PSF}, see \citealt{masseyOriginsWeakLensing2013,cropperDefiningWeakLensing2013} for an early description of how the \Euclid requirements were derived. This in turn requires a meticulous approach in data processing. This paper describes one of the earliest stages of the ground-segment pipeline: the set of processing elements that takes raw VIS images and produces scientifically validated and calibrated images on which photometric and shape measurements can be made. In particular, we describe the algorithmic choices we have made for each of the different instrumental effects that must be characterised and calibrated. We demonstrate the performance of each element individually where possible, as well as the measured performance on the first public release of data from the ground segment, the Quick Release 1 \citep{Q1-TP001,Q1cite}. 

\section{\label{sc:VISCamera} The VIS instrument}

In contrast to general-purpose observatories such as the \ac{HST}, which are designed for a range of different science cases, the ambitious \Euclid mission requirements described above drove the entire design of the VIS camera. The primary requirement is to have the best possible image quality over the largest possible field of view with minimal temporal variation. This in turn specifies the field of view of the VIS camera, the pixel scale, the exposure time for each observation, the survey duration and scheduling as well as many other aspects of the spacecraft \citep{EuclidSkyVIS}.   

\subsection{The VIS camera and detectors}
The VIS \ac{FPA} comprises 36 back-illuminated $40\,\micron$-thick \acp{CCD}, each $4132\times4096$ pixels in size. Each pixel is $12\,\micron\times12\,\micron$. At the centre of the \ac{FPA}, the pixel scale is $\sim \ang{;;0.1}$. The variation in intrinsic pixel size is less than 1\%. There are four readout nodes (with separate electronics), one at each corner of the \ac{CCD}. The e2v device CCD273-84 used was designed specifically for VIS. There are 51 serial prescan pixels, 29 serial overscan pixels, and 20 parallel overscan pixels. The prescan pixels are real pixels not exposed to light. The overscan pixels, whether serial or parallel, do not correspond to imaging pixels and are generated by overclocking the detector. These overscan regions enable the estimation of the pedestal level added by the detector electronics.

\begin{figure}[htbp!]
\centering
\includegraphics[width=1.0\hsize]{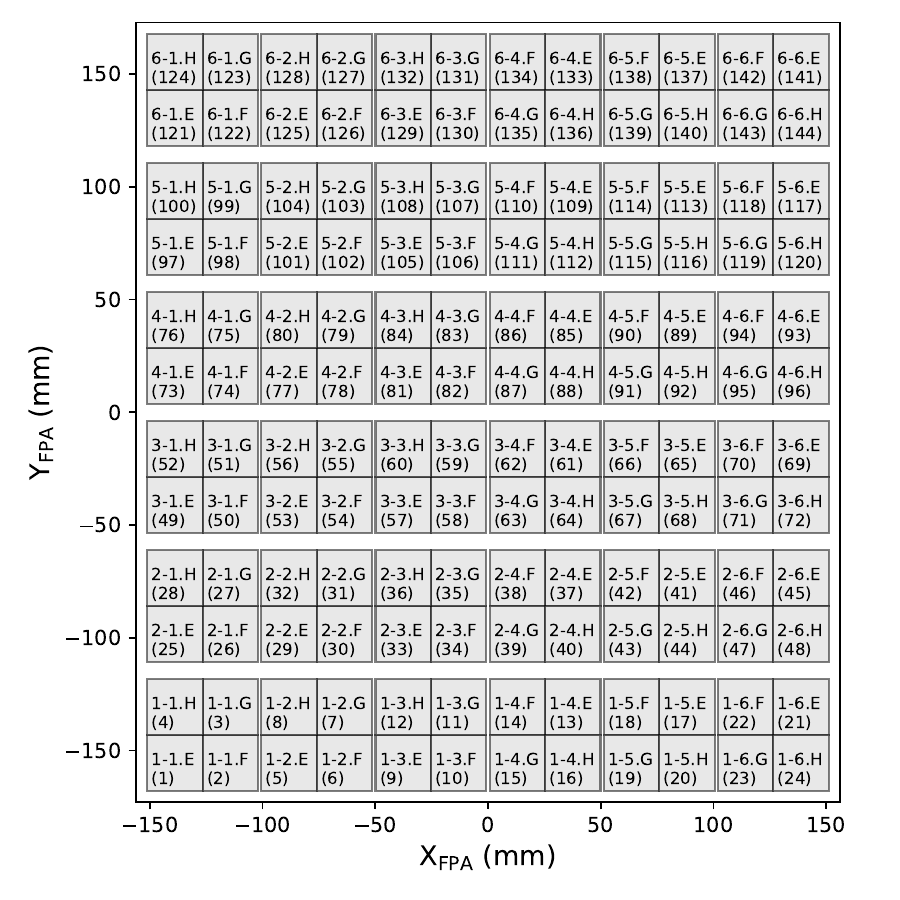}
\caption{Layout of quadrants in the \ac{FPA}. The detectors, each containing four quadrants, are arranged in a \(6\times6\) array. The mean \ac{CCD} spacing is \SI{1.53 \pm 0.03}{\milli\meter} and \SI{7.74 \pm 0.06}{\milli\meter} in the \(x\)- and \(y\)-directions, respectively.}

\label{fig:focal-plane-layout}
\end{figure}

The non-ionising effects of particle radiation on silicon detectors has been a significant concern since the mission was proposed. Energetic particles can produce defects in the silicon lattice that increase the \ac{CTI}. The VIS detectors contain a charge-injection structure in the middle of the device, allowing an arbitrary amount of charge to be introduced in a \ac{CCD}, which can then subsequently be read out (see figure~11 in \citeauthor{EuclidSkyVIS} \citeyear{EuclidSkyVIS}). The modelling and correction of \ac{CTI} using this charge-injection procedure is described in detail in \cref{sc:CTI}. The presence of the charge-injection lines means that a distortion model must be computed separately for each quadrant. For this reason, all calibrated images (or `frames') are delivered as a \ac{MEF} file with 144 extensions, one per quadrant. \Cref{fig:focal-plane-layout} shows the position of each quadrant in the \ac{FPA} in millimetres, labelled with its identifier and extension number. 

\section{\label{sc:PipeOver} The VIS processing function: Corrections made at the pixel level}

\begin{figure}[htbp!]
\centering
\includegraphics[width=1.0\hsize]{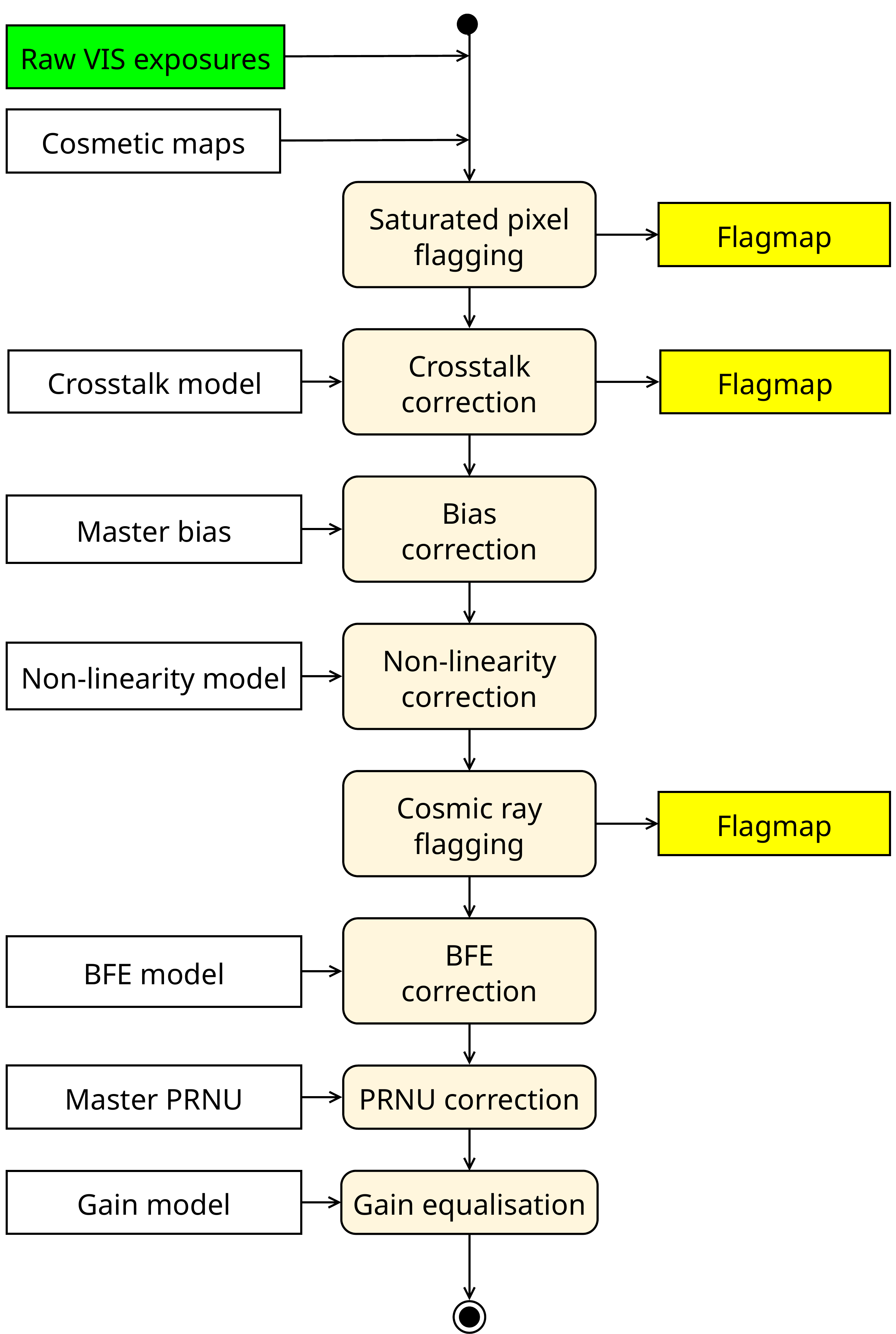}
\caption{First elements in the \ac{VIS PF}. Input calibration products are shown on the left and output flagmaps are on the right. `BFE' refers to the brighter-fatter effect 
(\cref{sc:BFE}), and `PRNU' is the pixel response non-uniformity (\cref{sc:flat}).  
We note that \ac{CTI} correction is not activated in Q1 processing due to limited radiation damage at this stage of the mission, and therefore it is not shown.}
\label{fig:quadrant-level}
\end{figure}

\subsection{\label{sc:intro} Overview}

The \ac{VIS PF} is divided into two main functional units: a `calibration pipeline' that computes the master calibration products for each of the instrumental effects and a `science pipeline' that applies each of these corrections to the raw VIS data. Here we describe the calibration and correction of each of these instrumental effects in the order they are applied in the science pipeline. Before launch, \ac{VIS PF} was developed, tested, and validated by processing highly realistic simulated raw VIS data \citep{EP-Serrano} created using the best available ground-calibration data. 

To understand the functionality required from \ac{VIS PF}, it is worth briefly describing the \ac{ROS},  described in \citet{Scaramella-EP1,EuclidSkyOverview} and executed at every location in the \ac{EWS}. Each \ac{ROS} consists of four dithers in an \,`S'\,-pattern, where VIS and NISP observe simultaneously. At each dither, VIS takes a 566\,s  `nominal science' exposure. The first two dithers also contain an additional 95\,s VIS `short science' exposure, and calibration exposures fill the remaining two dithers.\footnote{As a consequence of the shutter motion, the exposure times of  560.52\,s and 89.52\,s for long and short exposures (reported as \texttt{EXPTIME} in the image headers) corresponds to the time the shutter is fully open. In a given image, the total exposure per pixel varies over the field of view. See \cref{sc:ILL} for more details.} With this dither pattern, at least 95\% of the pixels are covered by at least three or four nominal exposures. While the \ac{EWS} comprises a single \ac{ROS} at each pointing, the \ac{EDS} fields are visited at least 40 times in a manner similar to the \ac{EWS} at a range of roll angles. The \ac{VIS PF} must therefore provide calibrations for both the nominal and short-science exposures.

We should note that the term `processing element' (PE) refers to a sub-element of the \ac{VIS PF}. Normally, the scope of a given processing element is limited. The term `processing function' refers to the ensemble of PEs that makes up the entire pipeline. In our case, \ac{VIS PF} refers to the sequence of PEs that comprise the VIS processing pipeline.\footnote{In \Euclid terminology, `OU-VIS' refers to the organisational unit that encompasses the VIS processing. It is responsible for defining the algorithms implemented in the \ac{VIS PF} and the validation of their outputs.} We note here that the principal client processing functions for \ac{VIS PF} are the shear processing function which measures shapes on individual VIS images, and the \ac{MER PF}, described in \citealt{Q1-TP004}, which combines data from \Euclid instruments with external data. \ac{MER PF} takes as input from \ac{VIS PF} the calibrated images described here and an associated \ac{PSF}  model. 

In operations, the main \ac{VIS PF} science pipeline (\texttt{VIS\_ProcessField}) is run on every raw science VIS exposure. It applies all the instrumental corrections in the order described below. This pipeline must be robust against bad or incomplete data, and it must also flag data so downstream PFs can apply selection functions if needed. Additionally, given the large number of VIS exposures arriving each day (around 156 during nominal wide-survey acquisitions), data must be processed quickly. This is because downstream PFs also have significant computational requirements, and the ground segment must be able to process all data from a given day's observations before the next day's data arrive. 

The \ac{VIS PF} is large and contains many processing elements. In the interests of simplicity, we divide the processing function into three sections.  In the first part, described in this section and shown schematically in \cref{fig:quadrant-level}, pixel-level corrections are carried out on the individual quadrants. In general, each quadrant is treated independently. Next, as described in \cref{sc:catalogue-level} and shown schematically in \cref{fig:catalogue-level}, catalogues are extracted from the processed quadrants and are used to compute the astrometric and photometric solutions. This stage provides the calibrated exposures that are used by SHE and MER processing functions. In the final stage, described in \cref{sc:stacking-level} and shown in \cref{fig:stacking-level}, we combine the individual calibrated frames from a \ac{ROS} to make a stacked image. 

\subsection{VIS LE1 processing}
\label{sc:LE1 processing}

Although not strictly part of the \ac{VIS PF}, for completeness we provide here a high-level summary of the software we developed to assemble the raw data from the VIS camera into \ac{MEF} files, the VIS 'Level 1' or LE1 processor. This software runs at the Science Operations Centre (SOC) and unpacks the compressed raw data transmitted from the spacecraft into a \ac{MEF} file with 144 extensions. The image metadata is processed from either the raw VIS data or other general housekeeping telemetry, and placed into the FITS headers with specific keywords.  The primary header of the \ac{MEF} contains information specific to the whole image, such as the observation date, image acquisition parameters, instrument temperatures, and spacecraft pointing and attitudes.  The header of each extension does not contain much information at this stage, apart from a few quadrant-specific parameters and the data compression ratio for the quadrant. It also contains a preliminary astrometric solution based on the commanded pointing values, RA, Dec, \ac{PA}, and the \ac{FPA} geometry (pixel size, dimensions, and layout of the detectors).

\subsection{\label{sc:BadPixelFlagging} Bad pixel flagging}
It is paramount for downstream processing functions that \ac{VIS PF} can correctly indicate the characteristics of each pixel in the individual calibrated frames. For this reason, each VIS calibrated image is accompanied by a flag (\texttt{FLG}) FITS file. Each pixel of this image is coded based on the processing during the \ac{VIS PF} ; a full description of the meaning of these flags is found in \cref{sc:q1-calframe-desc}. Below, we provide a brief overview of the primary artefacts identified in the images:  
 
\begin{itemize}

\item {\Ac{ADC} limit}. Data in VIS raw frames are integers produced by 16-bit \acp{ADC} that have an output range from 0 to 65\,535 \ac{ADU}. Any pixel at this maximum value is flagged as invalid because its actual value can be above this limit. 

\item {Saturated pixels}. Blooming can occur in thick \acp{CCD} (such as those used in VIS) when the charge is still below the full-well capacity of the \ac{CCD} creating a similar effect as saturation bleeding.  Above the blooming threshold, the charges in a pixel begin to overflow in the neighbour pixels in the same column. The value of this blooming threshold is different for each quadrant, and must be determined. The blooming-threshold calibration involves the computation of the \ac{PTC}, which is the relationship between the signal and its variance. Pairs of flat-fields at different fluences are subtracted to remove the lamp profile on the flat-fields, then the \ac{PTC} is created by computing the average signal and variance of the difference of the flat-fields in regions of $300\times300$ pixels. Once the \ac{BFE} and electronic nonlinearity are corrected (described later), we expect the relation to follow Poisson statistics, and therefore the variance equals the mean signal divided by the gain. The blooming threshold is then defined as the \ac{ADU} value above which the nonlinearity of the relation between signal and variance exceeds $5\%$. Blooming values range from 40\,kADU to 61\,kADU with a mean of 51\,kADU.

\item {Hot and dark pixels}. Abnormally bright or dark pixels are detected in master dark and flat frames as pixels with a mean value deviating by more than six or seven standard deviations from the master frame mean value. 

\item {Stitch-block boundaries.} This flag indicates the boundaries between different `stitch blocks'. The photolithographic  mask used in the manufacturing of the CCD273-84 exposes a block of $512\times256$ pixels at a time. Therefore, the mask is applied consecutively to the whole \ac{CCD} area, resulting in blocks that are stitched together. The positions of these blocks were identified from ground tests. 

\item {Bad cluster and columns}. In flat-field images, bad pixels are often surrounded by a cluster of brighter or darker than average pixels. This is true for both dust particles and electronic defects. Such clusters are detected by iteratively searching for a minimum of three nearest neighbours to a bad pixel or neighbour of a bad pixel, with a threshold of $\pm2$ standard deviations. An entire column is flagged as bad if it has more than 200 bad pixels in a master dark. 
 
\end{itemize}

\subsection{\label{sc:CT} Crosstalk measurement and correction}

In the VIS instrument, electronic crosstalk occurs because all twelve quadrants (four in each of the three \acp{CCD}) of the same \ac{ROE} are read synchronously in parallel. The signal in each quadrant receives a positive or negative contribution from potentially all other channels within the same \ac{ROE} through undesired capacitive or inductive coupling. This contribution is up to $5\times10^{-4}$ times the source quadrant signal. 

This effect is more significant when the electronic components of the quadrant channel are in proximity to another on the \ac{ROE} boards.  The value of the positive or negative coupling factor for each pair of source and victim quadrants is called the electronic crosstalk coefficient. There are 12 victim $\times$ 11 source quadrant pairs in each one of the 12 \ac{ROE} blocks, which amounts to a total of 1584 crosstalk coefficients that must be measured and corrected for in VIS images.

To measure a crosstalk coefficient, we select unsaturated pixels in a source and victim quadrant pair. Pixels in the source quadrant must contain a high signal (more than $10\,000$\,ADU) and the corresponding pixels in the victim quadrant and in the other quadrants of the \ac{ROE} must belong to the sky background. After background subtraction in the victim quadrant, the crosstalk coefficient is measured as the linear regression factor between the pixel values in the victim quadrant and the same pixel values in the source quadrant. Adding more exposures adds more source and victim pixel pair statistics and reduces measurement noise.

Once the 1584 crosstalk coefficients have been estimated, the electronic crosstalk correction in the science pipeline consists in subtracting, for a given victim quadrant image, the image of each source quadrant multiplied by the corresponding (positive or negative) crosstalk coefficient.

The largest crosstalk effect in a victim quadrant is around 30\,ADU from an almost-saturated source pixel. If a source pixel value saturates the \ac{ADC} (65\,535\,ADU), its `true' value that created the crosstalk effect is unknown, and all the corresponding victim pixels must therefore be invalidated.

The electronic crosstalk correction in the VIS science pipeline has two limits. First, to avoid mixing too many signals and adding noise or correlations, the crosstalk correction is not applied when the corresponding coefficient is under $4\times10^{-5}$ (a maximum effect of about 2.6\,ADU). Furthermore, we found that in addition to the linear effect currently measured and corrected, a differential effect also exists when sharp rising or falling edges occur in the source signal. This differential effect could amount to approximately 5\,ADU.

\subsection{\label{sc:Bias} Bias and dark}

The bias correction has two components. The first is the average pixel value of a quadrant relative to zero. This offset is expected to be a few thousand \acp{ADU} and is a single number per quadrant. This offset can be considered constant during the 72\,s readout, but has a long-term time dependency, and therefore it must be independently measured in every VIS exposure from the overscan pixels described above, which should not be contaminated by any signal source apart from readout noise. A second bias correction is made using a master bias image. This image captures the spatial, per-pixel dependence of the bias relative to the constant offset. The master bias image is computed from a median combination of at least 60 individual offset-corrected bias frames. These are zero-second, closed-shutter exposures. The bias is an additive offset, and therefore each exposure is bias-subtracted. 

Dark frames with 325\,s integration time are also taken, and subsequently combined into 
master-dark frames. \Acp{CR} detected in the individual dark frames are also used to produce a library of real \acp{CR} that can be used to generate survey images with simulated \acp{CR}, which is necessary to test the \ac{CR} detection software described in \cref{sc:CR}. 

Finally, shortly after launch, it was discovered that VIS images are affected in varying amounts by stray light (see Sect.~5.4 in \citeauthor{EuclidSkyOverview}). The master darks are used to assess the importance of stray light for a given set of exposures at a given solar aspect angle and telescope orientation.

\subsection{\label{sc:NL} Nonlinearity measurement and correction}

During the \ac{CCD} quadrant readout, the charge from all pixels passes through the quadrant's output node and readout electronics. However, because of limitations of the readout amplification chain, output digitised values may deviate slightly from perfect proportionality to the input charge. This effect, known as nonlinearity, may be as large as a few percent at high signal levels. Without correction, this would introduce an additional source of error in the measurement of the \ac{PSF}, as these measurements are carried out on bright sources with high signal-to-noise ratios. Simulations carried out before launch indicated that nonlinearity should be known to better than one part in $6\times10^{-4}$ to meet the requirements on \ac{PSF}  measurement at fluences above approximately $10$\,k\,e$^{-}$. 

Conceptually, correcting this nonlinearity effect is straightforward: measure the response in \acp{ADU} for a linear source of photons. Before launch, several algorithms were tested with simulated data to make this measurement. However, due to the presence of \acp{CR} and telescope jitter, these techniques could not deliver the percent-level accuracy required in real data. Work in perfecting these algorithms is still in progress. 

\begin{figure}[htbp!]
\centering
\includegraphics[width=1.0\hsize]{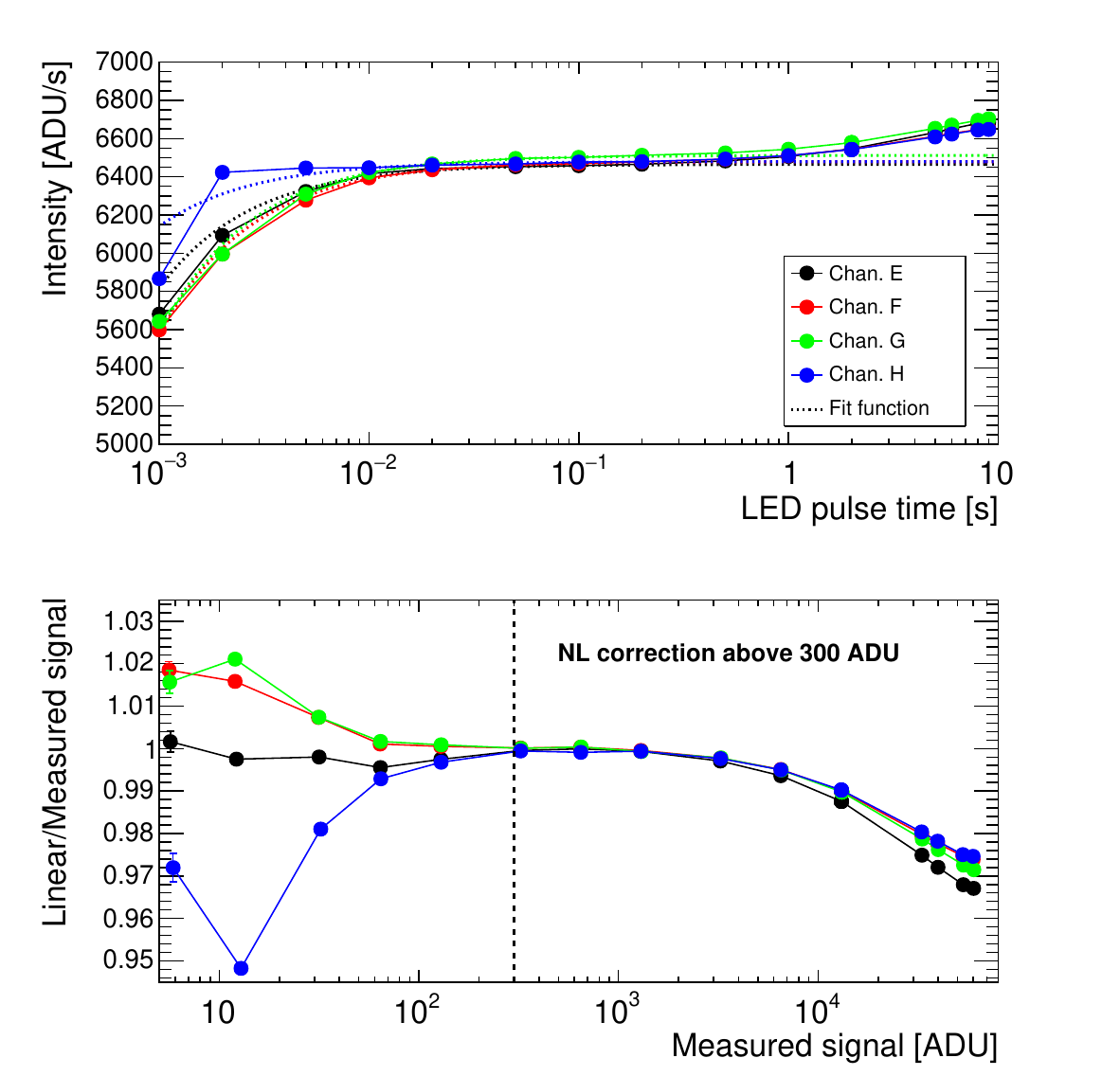}
\caption{Pre-launch measurements of VIS nonlinearity based on ground testing for \ac{CCD} 6-2. Each of the four lines corresponds to a separate quadrant. \textit{Top panel}: Measured intensity as a function of duration. \textit{Bottom panel}: Linear signal divided by measured signal.}
\label{fig:nonlinearity}
\end{figure}

However, during ground testing, we investigated an alternative technique that uses the \acp{LED} of the VIS calibration unit (CU). As described in \cref{sc:flat}, these diodes are used to illuminate the \ac{FPA} to calibrate the PRNU. The linearity and photometric stability of these had not been specified for the VIS. However,  ground testing in 2021 showed that over a one-hour timescale, the \ac{LED}  output varies by less than one part in $10^4$, indicating that this technique could be a viable method to correct for nonlinearity. Therefore, an experiment was designed to measure the nonlinearity during on ground testing by illuminating the \ac{FPA} with pulses of different length in a random order of calibration unit \ac{LED}  3 (720\,nm) and then reading out a windowed section of each detector. The nonlinearity of each channel was derived by determining the deviation of the signal from a linear function of the \ac{LED}  pulse time, assuming a constant \ac{LED}  intensity. Above 300\,ADU and until about 1000\,ADU, the signal is approximately linear. An example nonlinearity measurement is shown in \cref{fig:nonlinearity}. The nonlinearity increases gradually to about 2\%--4\% until blooming or saturation sets in. The statistical error of this measurement is below $10^{-4}$. The systematic error is driven by the thermal stability of the \ac{LED}  within a single exposure, and could amount to a few $\times10^{-4}$.

These nonlinearity measurements were computed at discrete points, as depicted in \cref{fig:nonlinearity}. When the correction is applied, the correction factor between these points is calculated using linear interpolation.  However, no correction is applied below 300\,ADU because of high uncertainties in the calibration measurements at these low fluxes. We assumed that the response is linear at these levels. 

\subsection{\label{sc:CR} Cosmic ray flagging}

The \texttt{LACosmic} algorithm \citep{2001PASP..113.1420V} is used in \ac{VIS PF} to flag \acp{CR}. However, the earliest \textit{Euclid} images showed many \acp{CR} that were only partially flagged. In nominal VIS images, about 1.6\% of pixels are flagged as \acp{CR}. The shapes of on-orbit \acp{CR} differ significantly from pre-launch expectations. In particular, there are many `fat' \acp{CR} that were initially only partially flagged. Post-launch, to assess the best possible parameter set to correctly flag \acp{CR} and to assess performance, we created a series of simulated images based on real \acp{CR} detected in VIS dark frames. From these, we can assess the performance of different algorithms using the receiver operating characteristic (ROC) curves, which assess how well a model distinguishes between two classes by plotting the \ac{TPR} against the \ac{FPR} for a range of settings. 

We follow the usual definition of the \ac{TPR}, which is the ratio of the number of \acp{CR} that are actually detected to the number of actual \acp{CR} in simulation. Conversely, the \ac{FPR} corresponds to the number of pixels that are incorrectly classified as \acp{CR}. The ratio is computed against the total number of pixels that are not actual \acp{CR} in the simulations. False positives (FP) are instances in which \acp{CR} are misclassified. An additional complication is that we must minimise the number of cores of non-saturated stars identified as \acp{CR}. After many simulations, we settled on the following set of parameters: $\texttt{sigclip}=10$; $\texttt{sigfrac}=0.4$; $\texttt{objlim}=20.0$; $\texttt{cleantype}=\texttt{medfilter}$. This results in a $\text{TPR}=78.3\%$ and an $\text{FPR}=0.03\%$, $\text{star FP}= 1.1\% \, (\IE < 20.0)$. 

\begin{figure}[htbp!]
\centering
\includegraphics[width=1.0\hsize]{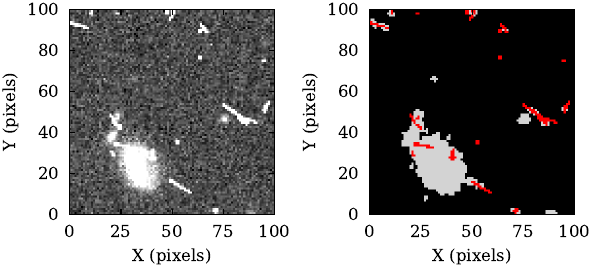}
\caption{Section of a VIS image (\textit{left panel}) and the part of the associated flagmap (\textit{right panel}). Grey pixels in the flagmap are objects, and red ones are identified as \acp{CR}.}
\label{fig:Q1CRflagging}
\end{figure}

If more than 5\% of pixels in a quadrant are flagged as \acp{CR}, all its pixels are flagged with the \ac{CR} region identifier,  \texttt{CR\_REGION}. This way, we can mask the `X-ray' pattern present on the VIS images during Solar flares (see Fig.~21 and Sect.~5.3 in \citealt {EuclidSkyOverview}). This enables other quadrants below the threshold to be used for scientific exploitation without having to discard the entire image.

\subsection{\label{sc:CTI} Charge-transfer inefficiency}

The performance of \ac{CCD} imaging sensors gradually degrades due to radiation damage. High-energy particles create defects in the silicon lattice that trap photoelectrons from one pixel and release them after a delay, characterised by an emission time, $\tau$. When this happens during readout, the electron is spuriously delayed into the wrong charge packet, appearing as if it were in a different pixel, and the image becomes smeared. Measurements from \Euclid data indicate that faint sources are  (fractionally) more affected than bright sources,  as a constant number of delayed electrons has a greater relative impact.  This effect is known as charge transfer inefficiency or CTI.

No correction for \ac{CTI} has been applied to Q1 data. However, we are monitoring \citep{2024SPIE13092E..0PS} the growing density ($\rho_{i}$) of several trap species with characteristic emission time ($\tau_{i}$). For a bright point source (with ${\rm S/N}=200$ following \citealt{2015MNRAS.453..561I}) that might be used for \ac{PSF}  measurement far from the readout register, our model indicates that the net effect of parallel \ac{CTI} at the end of May 2024 causes $\Delta F/F\approx-(3\pm 2)\times10^{-4}$ fractional flux loss, and $\Delta y\approx(2\pm 1)\times10^{-3}$\,pixels of spurious astrometric shift, where uncertainties are the standard deviation between the best-fit value in every \ac{CCD} quadrant. The damage in all quadrants is growing roughly linearly over time: with additional, abrupt damage during Solar \acp{CME} and greater damage in the corner of the \ac{FPA} that is also most exposed to radiation during \acp{CME}. As the \ac{CTI} `signal' accumulates, these measurements are becoming more precise, and \ac{CTI} correction will be activated for a future data release.

To monitor CTI, we used two techniques that both exploit a facility of the \ac{CCD} electronics to inject precisely known patterns of charge. 

The first method, \Ac{TP},  works by electronically injecting a flat image of charge into the full device, then shuffling it forward and backwards repeatedly, using a fixed phase time ($t_{\mathrm{ph}}$). If an electron is captured by a trap with $\tau\approx t_\mathrm{ph}$, the repeated shuffling will produce a dipole pattern, as charge is released into an adjacent pixel. The intensity of this dipole can be modelled as

\begin{equation} \label{eq:TP_equation}
    I(t_{\rm ph}) = N \, P \, \left[\exp\left(-\frac{t_{\rm ph}}{\tau}\right) - \exp\left(-\frac{2t_{\rm ph}}{\tau}\right)\right] ~,
\end{equation}
where $N$ is the number of shuffles and $P$ is a measure of the `efficiency' of the trap, that is, how likely it is to capture a nearby electron. Since $I$ is a function of $t_{\rm ph}$, a dipole curve can be built up by repeating the \ac{TP} process using different values of $t_{\rm ph}$. By fitting this curve with Eq.~\eqref{eq:TP_equation}, we measure $\tau$ and $P$ for every trap. Doing this for the full \ac{FPA} will thus provide information about the position and emission-time constants of the traps in the devices, supporting the \ac{CTI} correction efforts.

The second, \Ac{EPER}, works by injecting rectangular regions of charge with sharp edges. We fit the trailing of electrons after each charge-injection block into pixels that should contain zero electrons. We model \ac{EPER} trails (in both parallel and serial directions) using a non-linear model, {\tt arCTIc} \citep{2010MNRAS.401..371M,2014MNRAS.439..887M}. 
However, to first order, their profile $I(x)$ is roughly a sum of exponentials 
 
\begin{equation} 
I(x)=\sum_{i} \rho_{i} \left[ 1 - \exp{\left(-\frac{t_\mathrm{cl}}{\tau_i}\right)}\right] ~ \exp{\left(-\frac{t_\mathrm{cl}}{\tau_i}x\right)} ~,
\end{equation} where $x$ is the pixel number, and $t_\mathrm{cl}$ the readout clock speed in seconds per pixel.  
The prefactor in the bracket normalises the total number of trailed electrons to $\Sigma_{i} \rho_{i}$. First pixel response is the deficit of electrons in the leading edges of each charge injection block (in the opposite direction to the \ac{EPER} trails). These electrons are the first to encounter empty traps, and many are captured (to eventually be released in the \ac{EPER} trail).

\subsection{\label{sc:BFE} Brighter-fatter effect}

The \ac{BFE} refers to the redistribution of charge within pixels as charge is accumulated \citep{2006SPIE.6276E..09D,antilogus2014}. This is a consequence of the electric field of the charge; as more charge accumulates, pixel boundaries change. This effect depends nonlinearly on flux, and therefore causes the \ac{PSF}  to become larger for brighter stars. A secondary consequence of the \ac{BFE} is a slower-than-Poisson increase in variance with flux.

To correct this effect, we compute a kernel to apply to the data. This kernel represents the deflection field of the accumulated charge for one pixel, and is represented by an $11\times11$ matrix. Simulations have shown these dimensions give the correction for the \ac{BFE}. The \ac{BFE} kernel is the solution of the Poisson equation with the pixel covariance as the source term. To compute the pixel covariance, a series of pairs of flat-fields at high fluence with the same pointing are subtracted from each other, to remove the non-uniform illumination resulting from the calibration lamp profile. Since the shutter is open during flat-field exposures, the flat-field is immediately followed by a short science exposure to enable the masking of stars and other objects that are present in the flats. Finally, we  mask all\acp{CR} and any bad pixels. The pixel covariance must be adjusted using a zero correction, which involves modifying the central pixel value of the covariance matrix to ensure that the total sum of all covariance values equals zero. This adjustment is performed before solving the Poisson equation to compute the \ac{BFE} kernel. 

This \ac{BFE} kernel is used to correct the science image using an iterative flux-conserving approach. First, the image is converted to electrons and flagged pixels are `inpainted' (where their values are replaced by nearby good pixels) to exclude them from the correction. The image is then convolved with the kernel to compute a `deflection potential'. At each iteration, the algorithm calculates the required pixel-to-pixel flux transfers from the gradients of this potential. Flux is conserved by ensuring that any charge redistributed from one pixel is added to nearby pixels according to the kernel's deflection field. This creates a correction image that is then added to the original to redistribute charge based on the BFE kernel. This procedure is repeated until the absolute difference between successive iterations drops below  $10^{-6}$. The image is the converted back to ADUs and inpainted pixels restored to their original values.

\begin{figure}[htbp!]
\centering
\includegraphics[width=1\columnwidth]{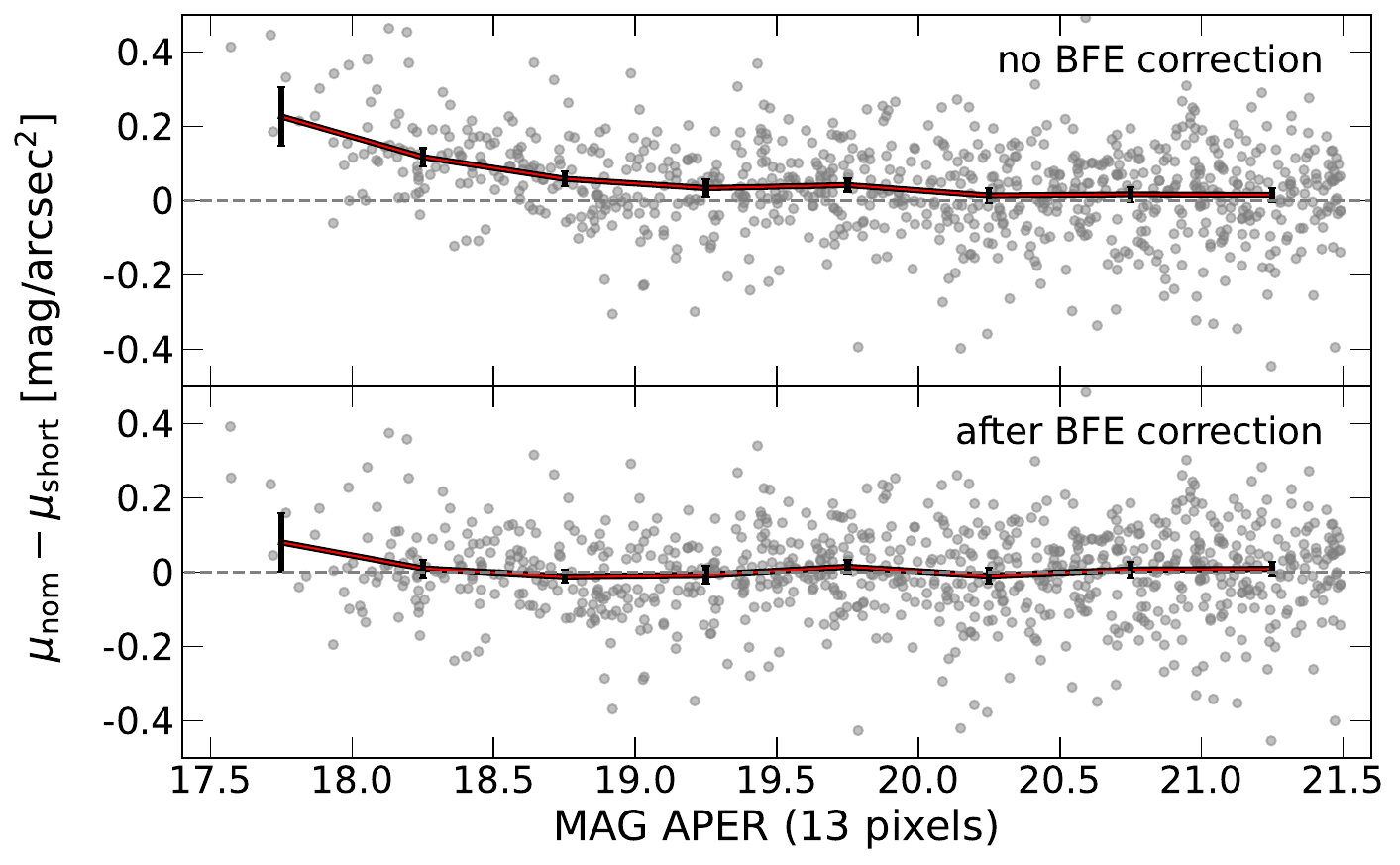}

\caption{Effect of \ac{BFE} correction on the difference of the surface brightness of the brightest pixel of stars in nominal versus short exposures as a function of VIS magnitude in an aperture of 13 pixels, or \ang{;;1.3} diameter.}
\label{fig:bfe-mumax}
\end{figure}

The photometric impact of the \ac{BFE} correction on stars is shown in \cref{fig:bfe-mumax}. Without \ac{BFE} correction, the \ac{PSF} core is broader for brighter stars. Consequently, in an aperture smaller than the scale at which the \ac{BFE} is important, the flux ratio of identical stars in short and nominal exposures is smaller for brighter stars, as shown on the top panel of \cref{fig:bfe-mumax} from the ratio of the flux of the brightest pixel. This effect is visible mainly for stars with $\IE<19$, and is not significant after \ac{BFE} correction (bottom panel).

\begin{figure}[htbp!]
\centering
\includegraphics[width=1\columnwidth]{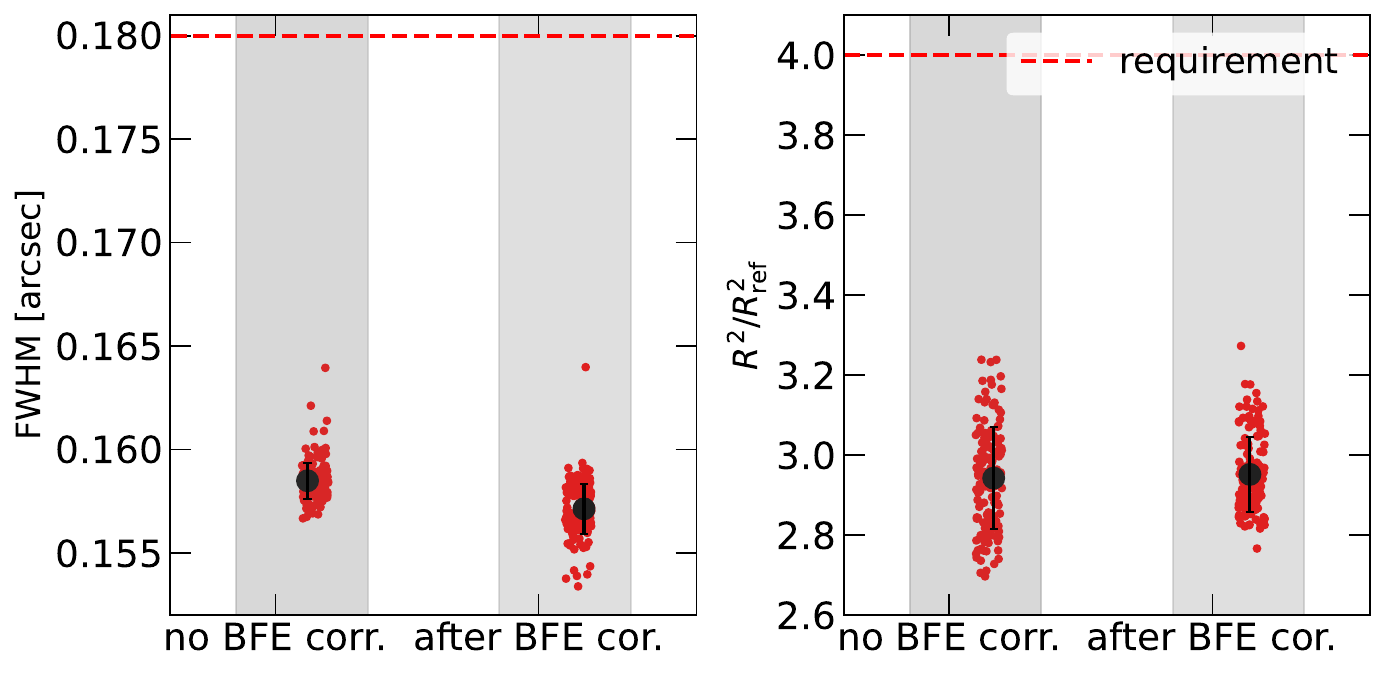}

\caption{Effect of \ac{BFE} correction on the \ac{PSF} model, as quantified from the \ac{FWHM} (\textit{left panel}) and $R^2/R^2_{\rm ref}$ (\textit{right panel}), which is the squared radius relative to that of a Gaussian profile with a \ac{FWHM} of \ang{;;0.2}. There is one red marker per quadrant. The black markers and error bars indicate the mean and standard deviation across all quadrants.}
\label{fig:bfe-before-after}
\end{figure}

\Cref{fig:bfe-before-after} shows the impact of \ac{BFE} correction on the \ac{PSF}  model derived with {\tt PSFex} \citep[][see also \cref{sc:PSF}]{2011ASPC..442..435B} in terms of the Gaussian \ac{FWHM} (left panel) and size $R^{2}$ in all quadrants before and after correction. (For a complete definition of the quantities used, see \cref{sc:image-notation}). The measured $R^{2}$ is compared to the reference value $R_{\mathrm{ref}}^{2}$ for a Gaussian with $\mathrm{FWHM}=0.20''$.

While this correction decreases the \ac{FWHM}, it leaves the \ac{PSF} size unchanged on average, since this metric provides information on the \ac{PSF} at scales larger than the \ac{BFE} correction. However, the correction still reduces the spatial variations on the \ac{FPA} (the standard deviation across all quadrants is smaller).

\subsection{\label{sc:flat} Pixel response non-uniformity}

The \ac{PRNU} is defined as the relative response of a pixel compared to the average of its neighbouring pixels. Larger pixels created by small variations in the photolithographic mask used for fabrication will have a larger response. The intrinsic and processed conditions of each pixel will also contribute to response differences. In \Euclid detectors, the amplitude of this pixel-to-pixel variation is around 1\%. The \ac{PRNU} correction (using a flat-field image) attempts to remove the differential pixel-to-pixel response. Because the \ac{PRNU} is both wavelength- and fluence-dependent, it is measured by illuminating the \ac{FPA} with \acp{LED} of four different wavelengths (573\,nm, 592\,nm, 638\,nm and 697\,nm) and three different fluences (10, 25, and 40\,kADU). During this procedure, the shutter must be open so that the \ac{LED} can illuminate the FPA, and consequently astronomical objects also appear. Therefore, for each fluence level, around 10 to 30 dithered exposures are taken. 

When producing each master \ac{PRNU} flat, each quadrant is examined for artefacts. All frames are corrected for crosstalk, bias, and nonlinearity. The shape of the illumination and the background is assumed to be stable to within the shot noise of the \ac{LED}  illumination. The master \ac{PRNU} flat at each fluence is produced from a clipped combination of the individual exposures and divided by the illumination profile (fitted as a spline surface). This procedure works well except at the 10\,kADU flats where some pixels from objects may remain. With this procedure, one master \ac{PRNU} flat is produced for each combination of \ac{LED}  and fluence. The final cosmetic flagmap is a logical `or' combination of all the individual master \ac{PRNU} flagmaps for all wavelengths and fluences. The final `small-scale' flat correction is the average of the four 
10\,kADU master \ac{PRNU} flats per wavelength. Finally, since the \ac{PRNU} is a multiplicative effect, it is corrected by dividing each image by this small-scale flat.

\subsection{\label{sc:bkg} Background estimation}
We compute the background for each VIS calibrated frame using the \texttt{NoiseChisel} program of GNU Astronomy Utilities (version 0.16; \citealt{2015ApJS..220....1A}) on each quadrant. By applying an initial threshold below the sky level and using mathematical morphology operators such as erosion and dilation, this can detect very faint signal in the images and provide a very accurate estimate of the background (defined as the mean of pixels outside sources). This is especially important to preserve the low surface brightness signal in the stacked images and not over-subtract them as background. One particular point that needed to be addressed when running \texttt{NoiseChisel} on \Euclid data is the extremely high density of \acp{CR} in some frames. To address this, all pixels flagged as \acp{CR} are set to not-a-number (NaN) before background estimation. \texttt{NoiseChisel} can ignore NaN-valued pixels during operation, which allows us to make a sufficiently accurate estimate of the sky background even when the \ac{CR} density was extremely high ($15\%$ of detected \ac{CR} pixels). 

 Since the \ac{EWS} continuously covers the sky, some VIS quadrants may see considerable signal, such as when imaging a large galaxy or cirrus cloud. This prevents \texttt{NoiseChisel} from estimating the background to the desired precision. In such cases, \texttt{NoiseChisel} is run a second time with less restrictive run-time options. If the signal is so strong that the quadrant contained insufficient noise, the \texttt{SExtractor} background map is used instead. Consequently, there is some  over-subtraction in the inner regions of objects that are larger than one quadrant. This problem will be addressed in future versions of \ac{VIS PF}. Fortunately, such bright objects are rare and for most of the sky this strategy of a multi-tier \texttt{NoiseChisel} background estimation produces high-quality background maps and stacked images.

\section{\label{sc:catalogue-level}The VIS processing function: Astrometric and photometric calibrations}

\subsection{\label{sc:Astro:intro} Introduction}
\begin{figure}[htbp!]
\centering
\includegraphics[width=1.0\hsize]{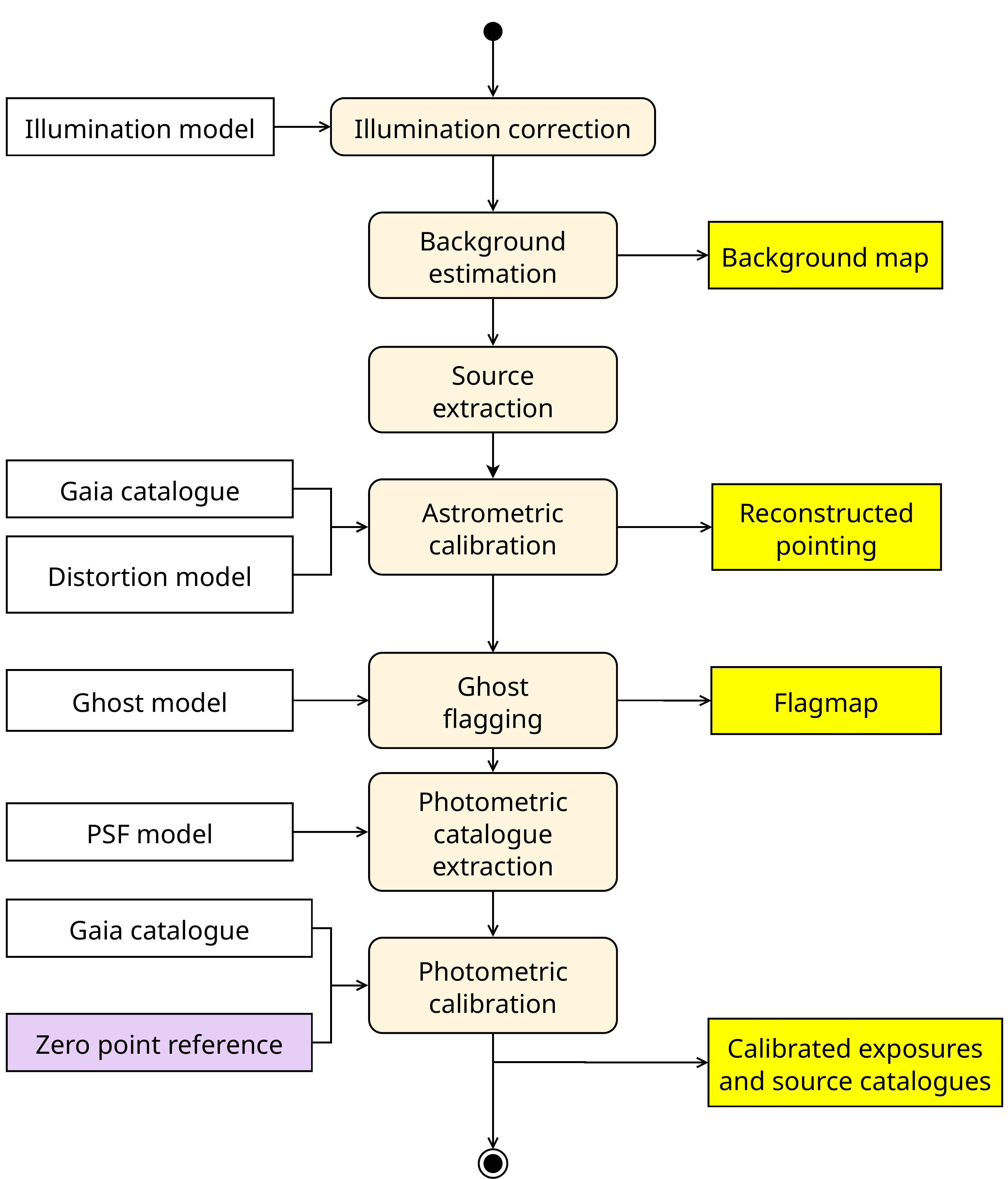}
\caption{Photometric and astrometric calibration and catalogue extraction. In this intermediate stage of the \ac{VIS PF}, many operations take place at the catalogue level. The output products from this stage correspond to the  data products that have been delivered for the Q1 release.}
\label{fig:catalogue-level}
\end{figure}

At the second level, quadrants are gathered and processed together, as illustrated in \cref{fig:catalogue-level}. Next, a series of calibrations are applied. The output products from this part of the processing chain comprise the Q1 data products.  

\subsection{\label{sc:gainequal} Gain equalisation}
In \ac{VIS PF}, we deliver images in \acp{ADU}. Each quadrant has a slightly different gain and so the final step is this part of the processing function is to apply a correction factor to equalise the gains. For Q1, the gain level is 3.48\,e$^{-}$\,ADU$^{-1}$. The correction factor is computed simply from the ratio of the individual detector gains and this reference gain. 

We determine the gains for each quadrant from a series of nine pairs of 1\,kADU flats. The difference images of each pair are combined with a clipping algorithm. Then, for each quadrant, we compute the gain as the average flux divided by half the variance of the pair difference in tiles of $300\times300$ pixels. The final gain is computed as an average over all tiles.

\subsection{\label{sc:PSF} Point spread function characterisation}

The highly detailed \ac{PSF} model required for \Euclid's shear measurements is provided by the shear-measurement processing function (also known as SHE-PF). In \ac{VIS PF}, however, we derive a simpler model for monitoring and internal quality control purposes, referred to as the VIS \ac{PSF} model. It does not consider the chromatic dependence of the PSF. This model is used by \ac{MER PF} for photometric measurements on VIS images. 

\begin{figure}
    \centering
    \includegraphics[width=0.5\textwidth]{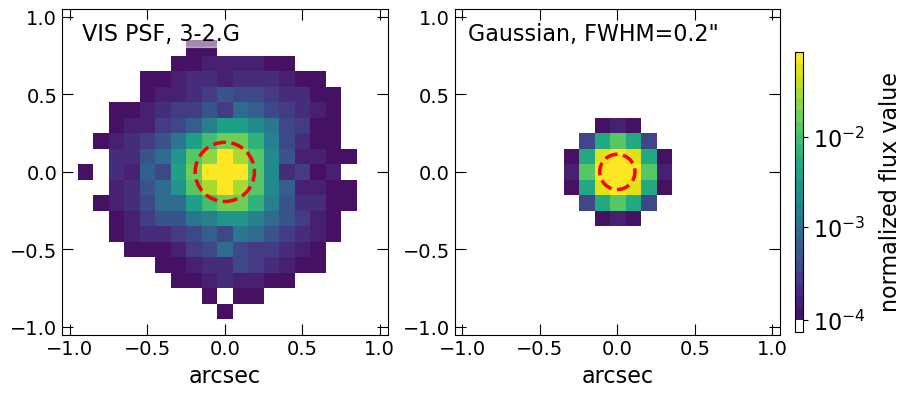}
    \caption{\ac{PSF} comparison. \textit{Left panel}: Q1 VIS \ac{PSF} model in one quadrant, measured in the self-calibration field with {\tt {PSFex}} at the VIS pixel scale. \textit{Right panel}: a Gaussian \ac{PSF} with a \ac{FWHM} of \ang{;;0.2}. The circles indicate the size of the PSF estimated using the moments-based method described in the text; in the left panel this is \ang{;;0.19} and the right panel it is \ang{;;0.11}.}
    \label{fig:psf_sampling}
\end{figure}

The VIS \ac{PSF} is estimated using {\tt PSFex} \citep{2011ASPC..442..435B}, which takes as input star positions and image stamps. The latter are 21 pixels wide, extracted with {\tt SExtractor} around the windowed star centroids on short exposures of the monthly visits of the self-calibration or `self-cal' field \citep{EuclidSkyOverview}. This field is located in the \Euclid Deep Field North at \ra{17;55;15}, \ang{+65;17;08} and has a sufficiently high stellar density to monitor a range of instrumental effects, including \ac{PSF} measurements. 

{\tt PSFex} optimally combines the input source images to enhance the  \ac{PSF}. More precisely, it iteratively computes the \ac{PSF} model based on the star images, compares the images to the model reconstructed at the star position, and excludes the detections that are too discrepant with respect to the model. The output of {\tt PSFex} is a \ac{PSF} vector that encapsulates the spatial variation of the \ac{PSF} across each quadrant, and in turn it allows us to reconstruct the \ac{PSF} model at any position. We assume here a second-order polynomial spatial variation of the PSF, modelled independently for each quadrant. \Cref{fig:psf_sampling} shows the VIS model \ac{PSF} in one quadrant compared to a Gaussian kernel with a \ac{FWHM} of \ang{;;0.2}. 

\begin{figure}
    \centering
    \includegraphics[width=0.48\textwidth]{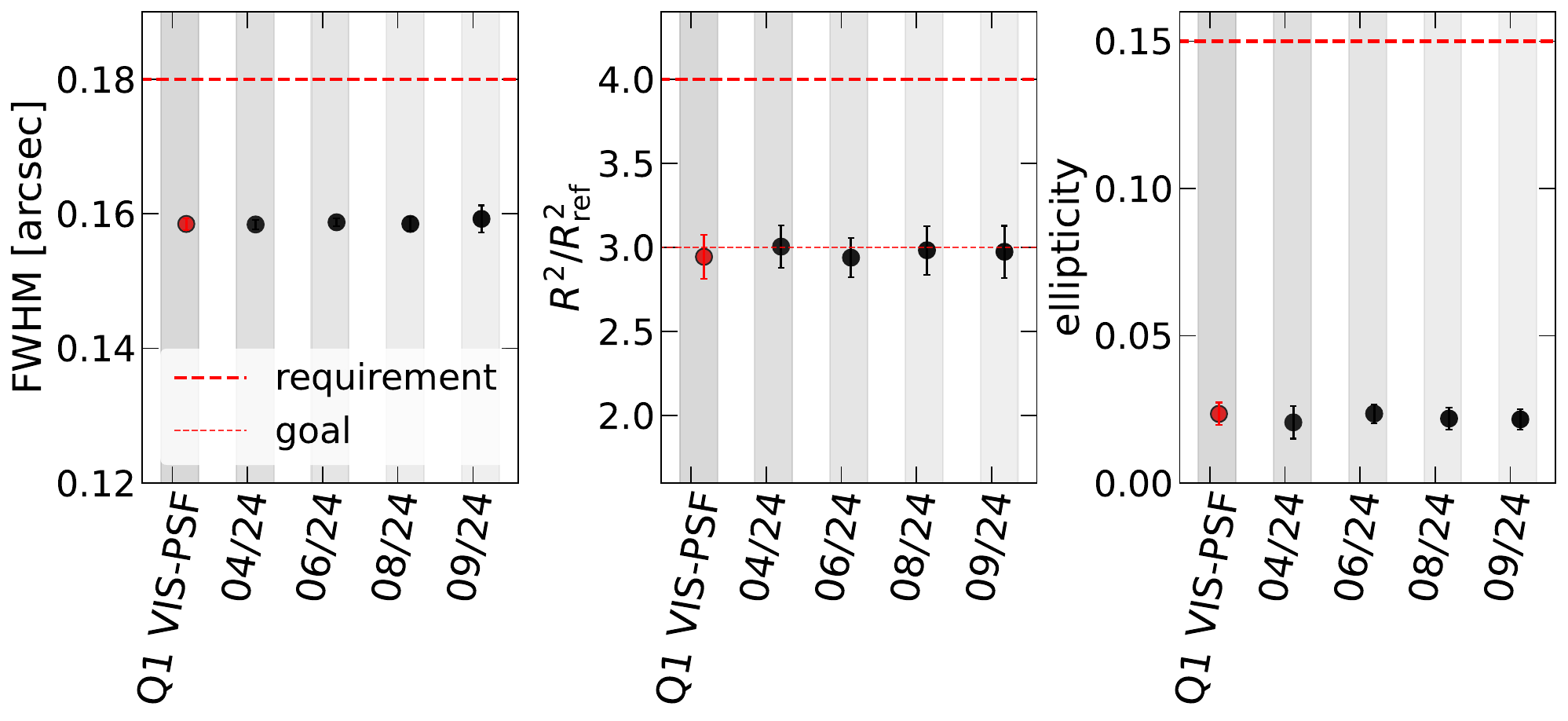}
    \caption{Stability of \ac{FWHM}, size, and ellipticity of the VIS \ac{PSF} model measured in the self-cal field. The red marker indicates the measurement on the Q1 VIS \ac{PSF} model. The values are measured in each quadrant, and then averaged, and the error bars indicate the dispersion of the measurements across the quadrants. The thick red dashed lines correspond to the mission \ac{PSF} requirements \citep{MRD:13}. Regarding the PSF size, \citet{MRD:13} also specifies a ``goal".}
    \label{fig:timevariations}
\end{figure}

We computed the size, ellipticity, and \ac{FWHM} of the model \ac{PSF} to monitor its  evolution with time and its variations across the \ac{FPA}, following the expressions in \cref{sc:image-notation}. First, the \ac{PSF} centroid is determined iteratively, after weighting the \ac{PSF} stamp with a Gaussian function having $\sigma=\ang{;;0.75}$. The \ac{FWHM} is calculated as the azimuthal average of the width of the \ac{PSF} at half maximum. As with the squared radius $R^2$ defined in \cref{eq:R2}, the ellipticity $e$ is measured using the second-moment method, with the same Gaussian weighting of  $\sigma=\ang{;;0.75}$ as described in \cref{eq:ellip_tot}.  
Each month, a series of observations are made in the self-cal field. Here, a carefully chosen set of dither patterns ensures that each source is observed at various locations across the \ac{FPA}. 

\begin{figure}
    \centering
    \includegraphics[width=0.5\textwidth]{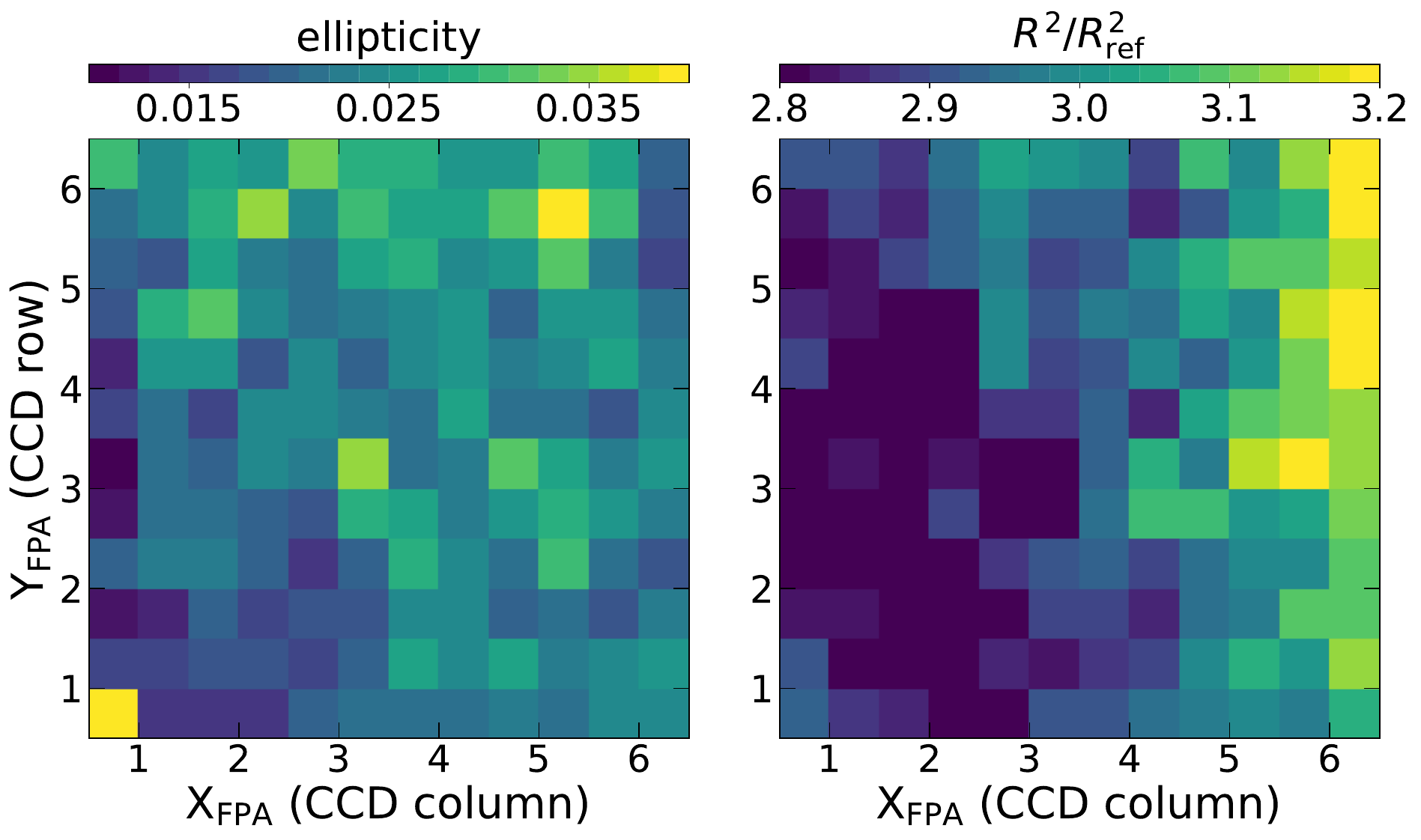}
    \caption{Variations of ellipticity (\textit{left panel}) and size (\textit{right panel}) of the VIS \ac{PSF}  model across the \ac{FPA}.}
    \label{fig:psf_fpa}
\end{figure}

The Q1 \ac{PSF} model has been calibrated on data from March~2024 self-cal observations, immediately after the first decontamination campaign. \Cref{fig:timevariations} shows the time variations of these metrics and compares them to the top-level mission requirement of \Euclid (red dashed line).  At the level of accuracy needed in \ac{VIS PF}, these metrics are very stable, and meet the mission requirements.  The \ac{FWHM} and ellipticity are comparable to those estimated with {\tt PSFex} in the Early Release Observations~\citep[see e.g.,][]{EROPerseusOverview,EROGalGCs}. In particular, the \ac{FWHM} of the Q1 \ac{PSF} model varies from \ang{;;0.157} to \ang{;;0.164} depending on the quadrant, with a mean over all quadrants at  \ang{;;0.158} and standard deviation of \ang{;;0.001}. We note that due to undersampling, the \ac{PSF} model estimated with {\tt PSFex} is generally slightly larger and rounder. \Cref{fig:psf_fpa} shows the variations of size and ellipticities over the \ac{FPA}. These spatial variations are relatively stable in time.

\subsection{\label{sc:ILL} Illumination correction}

In successive exposures, an identical object placed at different locations across the VIS \ac{FPA} will not have the same measured flux. The main cause of this effect is that the VIS shutter unit takes 2.7\,s to open and close, which leads to a left-to-right gradient in measured fluxes. Because the shutter motion time is fixed, this means there is a roughly 6\% left-to-right flux difference in short exposures and 1\% flux difference for nominal exposures. Other factors leading to throughput variations across the \ac{FPA} can include variations in optical elements such as mirrors, quantum efficiency variations between and within detectors, and errors in the gain estimation for each quadrant. The overall error in the uncorrected magnitudes is on the order of 0.05\,mag in nominal exposures. This implementation described here is based on code and algorithms developed for the near-infrared processing function described in \cite{Q1-TP003}. 

This large-scale flat-field is derived from regular observations of the self-cal field (\cref{sc:PSF}) by minimising the difference between the measured magnitudes of stars at all positions for 15\,000 unique stars. Because the amplitude of the effect is small, creating the purest sample of unsaturated stars is essential. The large-scale flat pipeline takes as input \texttt{SExtractor} reference catalogues constructed using a \ac{PSF} generated by running \texttt{PSFex} on the self-cal field (for the Q1 observations). We used the following selection criteria:

\begin{equation}
    \begin{aligned}
        \mathrm{\texttt{spread\_model}} & \quad |\mathrm{SPREAD\_MODEL}| < 0.005\,; \\
        \mathrm{Ellipticity} & \quad \mathrm{ELLIPTICITY} < 0.05\,; \\
        \mathrm{Flags} & \quad \mathrm{FLAGS} == 0\,; \\
        \mathrm{Signal\text{-}to\text{-}Noise} & \quad \mathrm{S/N} > 80\,.
    \end{aligned}
\end{equation}

The \texttt{spread-model} parameter (described in  Section~3.2.7 in \citealt{2012ApJ...757...83D}) measures the amount by which the light profile of an object differs from a point source and can be used to robustly select point-like sources, see \cref{fig:Spread-Model-Image}. 

The large-scale flat characterises the relative illumination in a coarse grid of $3\times3$ tiles per quadrant. It is derived separately for short nominal exposures to account for the shutter effect on the effective exposure time. The final product of this calibration is a coarsely binned map of the \ac{FPA} that contains a multiplicative correction that can be applied to all images, known as the `large-scale' flat.

\begin{figure*}[htbp]
    \centering
    \includegraphics[width=0.48\textwidth]{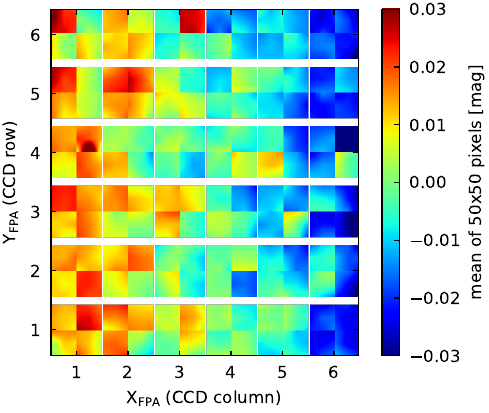}
    \hfill
    \includegraphics[width=0.48\textwidth]{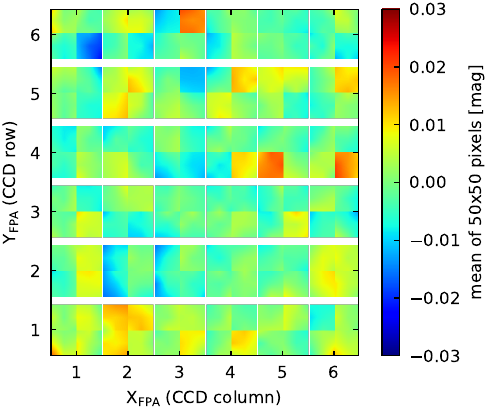}
    \caption{VIS large-scale flat for short exposures (\textit{left panel}) and nominal exposures (\textit{right panel}). The detector quadrants are easily discernable. This large-scale flat is computed by combining three successive visits of the self-cal field. For short exposures, the shutter effect on the measured signal is evident. The large-scale flat also corrects differences in the quantum efficiency and residual inaccuracies in the gain calibration, clearly visible for some quadrants.}
    \label{fig:largeflatcal}
\end{figure*}

In \cref{fig:largeflatcal} we show the large-scale flat correction derived for nominal and short exposures estimated from a combination of three successive visits of the self-cal field (taken on 10th June, 18th June, and 19th July 2024, respectively). In the left panel, the effect of the shutter movement on the measured signal is evident. For the nominal exposures in the right panel, the effect is less pronounced, as the ratio of the shutter movement time to the exposure time is correspondingly smaller and partially cancelled by the illumination in \Euclid's off-axis optical design \citep{EuclidSkyOverview}.

\subsection{\label{sc:AST} Astrometric calibration}

The astrometric calibration begins with extracting a Gaia DR3 reference catalogue \citep{2023A&A...674A...1G}, which overlaps every VIS exposure footprint. Gaia sources with five- or six-parameter astrometric fits are moved to the epoch of each VIS exposure, considering their proper motions and parallaxes. In addition, stellar aberration is incorporated using the spacecraft's orbital data added to the image headers by the ESA's mission operations centre. VIS catalogues are extracted from each image using \texttt{SExtractor} and bright, unsaturated stars are selected with $18.5<\IE<22$ and $16.7<\IE<22$ for nominal and short exposures, respectively. 

To characterise the detector layout and distortion, an approximate initial \ac{WCS} is available in the image header. This initial model is just the layout of the quadrants in the FPA, derived from the `as specified' values in the mission database. It does not contain any rotation or distortion.  This is typically sufficient to map pixel coordinates to equatorial ones to within a few arcseconds. This initial \ac{WCS} means a positional cross-match between the VIS catalogue and the reference star catalogue is relatively straightforward.  

Next, an initial model of the focal-plane layout and distortion (`\ac{FPA} model') is fitted using a least squares algorithm. From here, an iterative process is adopted, involving rematching between the VIS and reference catalogues (with stricter requirements), leading to an improved \ac{FPA} model. This process is repeated  until the required astrometric precision is attained, the reference source list does not change, or a set number of iterations is reached.

An \ac{FPA} model is a nearly complete \ac{WCS}, since it comprises reference pixel locations, a linear distortion matrix (\textsc{CDi\_j}), and Simple Imaging Polynomial (SIP; \citealt{2005ASPC..347..491S}) distortion coefficients. The only difference with a standard \ac{WCS} is that we use the individual quadrant centres for the reference-pixel coordinates instead of the centre of the \ac{FPA}. This choice avoids large positional offsets and minimises the reduction in SIP model sensitivity that would occur if the \ac{FPA} centre were used.

It is  desirable to characterise quadrant distortion separately from \ac{WCS} fitting, since not all exposures are expected to contain sufficient astrometric reference stars to measure the distortion with the required precision. SIP was adopted since it models distortion as a function of image coordinates, which are static with respect to the expected sources of distortion. This contrasts with TPV,\footnote{\url{https://fits.gsfc.nasa.gov/registry/tpvwcs/tpv.html}} which models distortion as a function of intermediate world coordinates, which is typically better suited to ground-based observations.

Production of the \ac{WCS} for a given VIS exposure requires a previously computed \ac{FPA} model that provides the basic quadrant layout and distortion model in the spacecraft-centric reference frame. The preliminary spacecraft pointing position and \ac{PA} are used to produce an initial \ac{WCS} from the \ac{FPA} model, and these are used to generate preliminary sky positions for the detected VIS sources. These are then cross-matched to the reference catalogue, and the pointing and \ac{PA} are then fitted using the least squares method. The process of cross-matching and pointing and position-angle refitting is repeated until either the required astrometric precision is attained, the reference source pool is static, or a set number of iterations is reached.

This procedure produces a \ac{WCS} that maps detector quadrant coordinates into equatorial coordinates in the spacecraft-centric reference frame. The conversion to a \ac{WCS} that maps to International Celestial Reference Frame (ICRF) coordinates is performed numerically using the spacecraft geocentric position and velocity. In the last step, the SIP \ac{WCS} is converted to the more commonly used TPV convention and these \ac{WCS} keywords are written to each of the 144 image headers (one per quadrant) in the calibrated frames. 

\begin{figure}[htbp!]
\centering
\includegraphics[width=1.0\hsize]{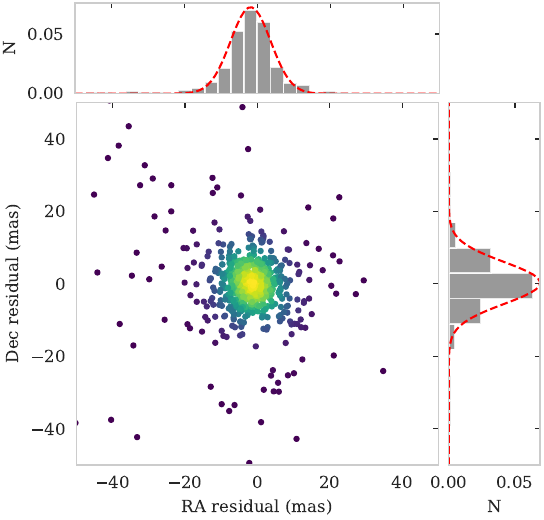}

\caption{Astrometric residuals in a single image of the Q1 release. We show the difference between the positions of Gaia sources propagated to the \Euclid\ observation epoch. Histograms of the RA and Dec residuals are overlaid with Gaussian fits (dashed red lines). The half-widths of the central 68\% intervals are $5.7$\,mas (RA) and $6.0$\,mas (Dec). The colours represent the source density. There are 709 sources in total.}

\label{fig:astrometry-extern-one-image}
\end{figure}

\Cref{fig:astrometry-extern-one-image} shows the difference between sources from the Gaia DR3 astrometric catalogue propagated to the \Euclid Q1 observation epoch and sources on one of the Q1 images. The width of the distributions -- measured as the median absolute deviation -- is approximately 5.5\,mas and 5.7\,mas for RA and Dec, respectively. 

\subsection{\label{sc:PHO} Photometric calibration}

Photometric calibration produces  calibrated magnitudes from instrumental magnitudes using reference sources. The initial absolute VIS photometric zero-point calibration is computed based on 80 repeated measurements of the primary white dwarf spectrophotometric standard WDJ175318.65+644502.15 (Gaia SourceID 1440758225532172032) in short exposures in the self-cal field. On each image, 13-pixel diameter instrumental magnitudes are measured on short exposures. A curve-of-growth analysis is used to compute the correction to total magnitudes. 

WDJ175318+6445 is part of a set of faint hot white dwarfs that were observed both photometrically and spectroscopically with WFC3/IR and STIS (Cycle 29 and 31) on the \ac{HST}. By combining GALEX, Gaia, and the \ac{HST} WFC3/IR data, the spectral energy distributions (SEDs) were fit with a grid of hot white dwarf templates. The best-fitting template was used to provide preliminary (1\,\text{--}\,2\% precision) synthetic VIS (and NISP) magnitudes. These SEDs will be refined over the coming months by combining the previous \ac{HST} data with new STIS observations of the stars. More details can be found in Appleton et al. (in prep.) and Deustua et al. (in prep.). Over the course of the \Euclid  mission, it is expected that more calibration sources will be observed to further improve the accuracy of \Euclid's  absolute calibration, which is directly linked to the \ac{HST} CALSPEC database \citep{bohlin2017}.

Soon after launch, it was realised (from comparisons of sources in the self-cal field observed at successive dates) that the telescope throughput was steadily decreasing due to the build-up of water ice on the mirrors in \Euclid's instrument cavity. This was expected and is common in spacecraft \citep{Schirmer-EP29}. A first thermal decontamination campaign was carried out on 12 March 2024 when folding mirror 3 (FoM3) and the tertiary mirror (M3) were heated to 160\,K \citep[for the telescopic design see][]{EuclidSkyOverview}. This restored the throughput to levels observed immediately after launch. However, the throughput rapidly dropped once more. After a second decontamination campaign on 6 June 2024, heating FOM3 only, the throughput returned to immediate post-launch levels and has remained stable since. 

These variations in observed throughput indicate that a single photometric zero-point is inadequate, especially when throughput changes rapidly, such as between the two decontamination campaigns. We therefore developed an alternative method to compute the zero-point for each exposure using a direct comparison with Gaia photometry. 

We take \( G \) as the Gaia \( G \)-band magnitude, and \( G_{\mathrm{BP}} \) and \( G_{\mathrm{RP}} \) are the Gaia blue and red magnitudes, respectively. The colour term \( C \) is defined as
\begin{equation}
    C = G_{\mathrm{BP}} - G_{\mathrm{RP}} - 0.3406\,,
\end{equation}
where the constant term transforms to the AB magnitude system \citep{oke1983}.
The VIS \( \IE \) magnitude is then
\begin{equation}
    \IE = G + 0.1136 - \left( p_0 + p_1 \, C + p_2 \, C^2 + p_3 \, C^3 + p_4 \, C^4 \right),
\end{equation}
with
\begin{align*}
    p_0 & = -0.00489071, \\
    p_1 & = 0.405557, \\
    p_2 & = -0.0256434, \\
    p_3 & = -0.0627437, \\
    p_4 & = 0.0204569.
\end{align*}

These coefficients were derived from synthetic photometry using the measured VIS throughput (Appendix~\ref{sc:VISCamera}) and the Gaia filter response functions. The transformation is valid for sources with intrinsic Gaia colours in the range \( -1.0 < G_{\mathrm{BP}} - G_{\mathrm{RP}} < 2.5 \). We note that  that considerable intrinsic scatter exists in this relation for very blue and red sources, partially caused by the very wide Gaia and VIS passbands that marginalise over many spectral types.

\begin{figure*}[htbp]
    \centering
    \includegraphics[width=0.9\textwidth]{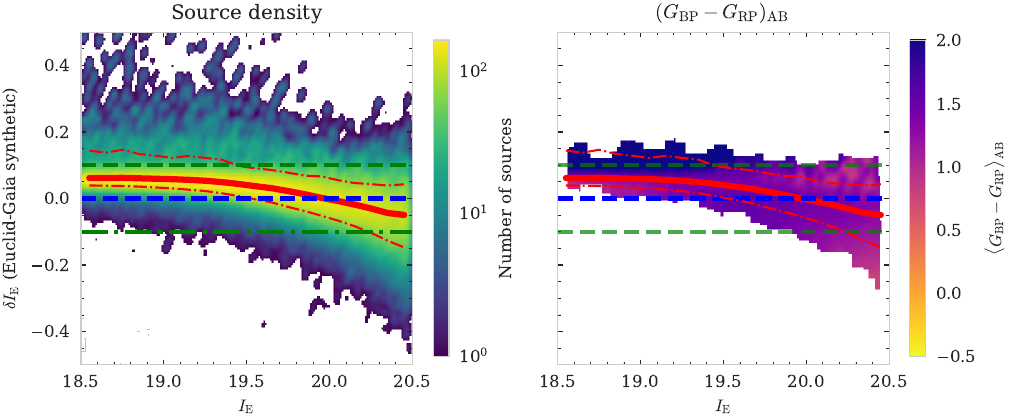}
    \hfill
    \caption{Difference between observed \Euclid $I_\mathrm{E}$ magnitudes and Gaia synthetic $I_\mathrm{E}$ magnitudes as a function of $I_\mathrm{E}$ for unsaturated stars in nominal exposures in the Q1 dataset. \textit{Left panel:} Stellar source density in the $I_\mathrm{E}$ vs.\ magnitude difference plane. \textit{Right panel:} Average $\langle G_{\mathrm{BP}} - G_{\mathrm{RP}} \rangle$ in AB magnitudes in the same parameter space. In both panels, the red line traces the median offset, and the dotted red lines show the region containing  68\% of the points. In both panels, the dashed horizontal lines at $\pm0.1$\,mag provide reference values.}
    \label{fig:gaiacomp}
\end{figure*}

These Gaia-derived zero-points allow us to monitor the VIS throughput and compute zero-points even for data that are affected by ice.  For Q1, the photometric calibration was computed by calculating the median of all individual zero-point measurements from July 1st to September 1st 2024. This value is 24.57\,ADU\,s$^{-1}$. 

For nominal and short exposures respectively, we select sources with $18.5<\IE<19.8$ and $16.5<\IE<19.8$, where the bright limit corresponds to the saturation limit for point sources and the faint limit it set by the faint-end photometric accuracy of Gaia. We use the \texttt{spread-model} parameter to select point-like source as before \cref{sc:ILL}. For each source, object photometry is performed using 13 pixel diameter circular apertures (corresponding to \ang{;;1.3}). We apply a correction factor to these magnitudes to arrive at `total' magnitudes which contains most of the object flux (the correction factor is computed by comparing fluxes in 13-pixel and 50-pixel apertures for non-blended stars over the \ac{FPA}; it is available in the header of each image in the \texttt{APERCOR} keyword). The zero-point is computed from a clipped estimate of the median difference between these VIS instrumental magnitudes and the predicted Gaia VIS magnitudes for Gaia stars with $0 < C < 2.5$.  

The left panel of \cref{fig:gaiacomp} shows the difference between VIS magnitudes for stars in nominal exposures in Q1 and synthetic VIS magnitudes derived from Gaia fluxes of the same stars. Despite the wide VIS bandpass, agreement is generally excellent. The small offset we find is a consequence of the aperture correction and suboptimal background subtraction in our Q1 standard star processing;  this offset has been reduced in subsequent processing. The right panel shows that for Q1 data $\langle G_{\mathrm{BP}} - G_{\mathrm{RP}}\rangle$ colours are approximately constant as a function of magnitude. 

\subsection{\label{sc:ghost}Ghost flagging}

\begin{figure}[htbp!]
\centering
\includegraphics[width=1\hsize]{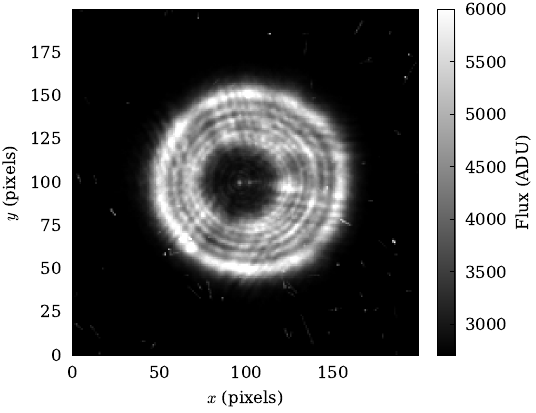}
\caption{Example of a VIS optical ghost from a bright star with \IE=5.5. The flux in the ghost pixels is around 3000--4000 \acp{ADU} above the background.}
\label{fig:ghost}
\end{figure}

The dichroic element \citep{EuclidSkyOverview} separates visible and infrared light in \Euclid. The nominal light path of VIS is reflected at the dichroic's front surface. A small fraction of the photons, however, is reflected at the back surface. The additional optical path length leads to out-of-focus ghost images of bright stars ($\IE < 14$) in the VIS focal plane. The attenuation factor is about $10^{-6}$ over the ghost area. The ghosts are offset a few hundred pixels from the star's position. An example ghost image is shown in \cref{fig:ghost}. 

A model of the ghost offset as a function of its source star position in the \ac{FPA} has been built using in-flight science images. It predicts the position and brightness of the dichroic ghosts. No attempt is made to model and subtract the ghosts themselves, as their surface-brightness distribution is rather complicated and chromatic. Pixels affected by a ghost are included in the image's flag map based on a simple flux threshold. We note that this step can only be carried out after the astrometric solution is made.

\section{\label{sc:stacking-level}The VIS processing function: Stacking and catalogue extraction}

\begin{figure}[htbp!]
\centering
\includegraphics[width=1.0\hsize]{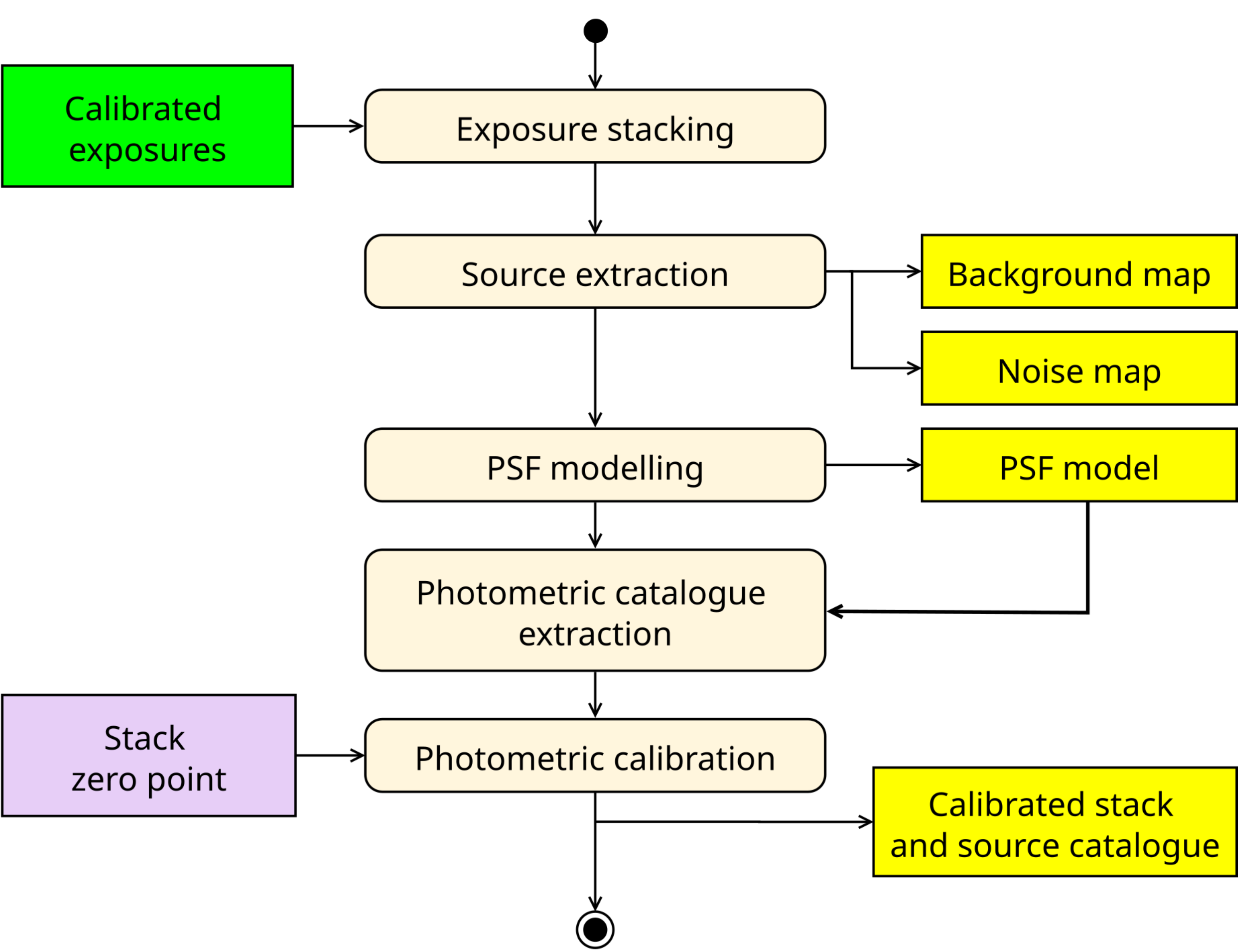}
\caption{Final processing elements of \ac{VIS PF} that produces stacked images and catalogues.}
\label{fig:stacking-level}
\end{figure}

\Cref{fig:stacking-level} shows the final stage of the \ac{VIS PF}. Stacked images and catalogues derived from these stacked images are not delivered as part of Q1, but they are described here in the interest of completeness.

\subsection{\label{sc:STA} Stacking}

\begin{figure*}[htbp]
    \centering
    \includegraphics[width=0.98\textwidth]{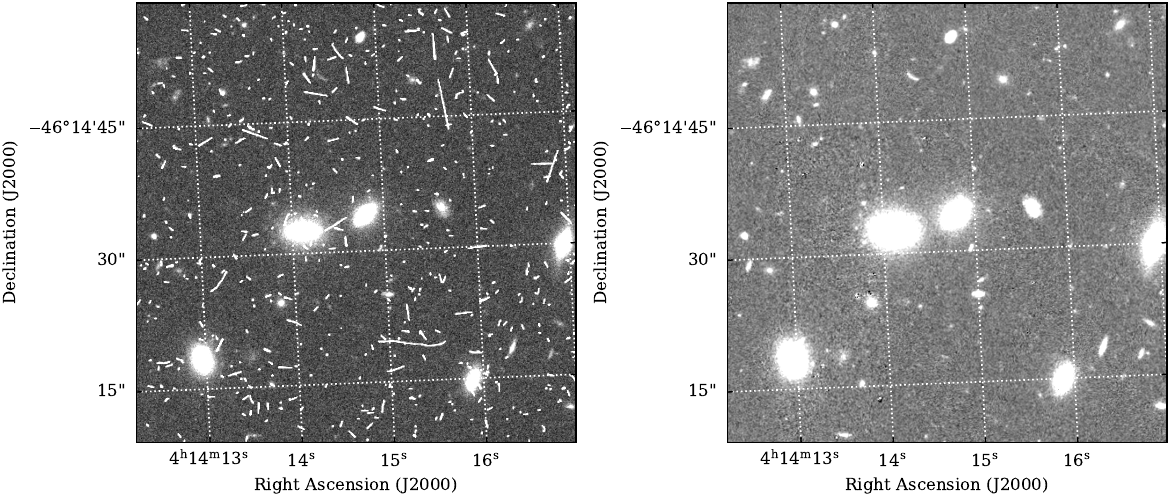}
    \caption{Part of VIS images in the Q1 data. \textit{Left panel}: Individual VIS calibrated frame. \textit{Right panel}: Stacked image of the same sky region, comprising six combined VIS images (four nominal and two short) for a total exposure time per pixel of 2422\,s. }
    \hfill
    \label{fig:images}
\end{figure*}

In the final processing steps, individual VIS calibrated frames and their associated weight maps are combined using the image resampling and co-addition software \texttt{SWarp} \citep{2002ASPC..281..228B}. Bad pixels in the weight maps are set to zero using the associated flag maps. \Cref{fig:images} shows a single VIS image (left panel) and a stacked image, combined from six VIS images, four nominal exposures and two short exposures (resulting in a total exposure time per pixel of 2422\,s). This stacked image has a $10\sigma$ depth of $\IE=25.6$, computed as the variance of \ang{;;1.3} diameter apertures, chosen to represent extended objects. This compares very favourably with the top-level \Euclid requirement of $\IE=24.5$. Together with the stacked image, an inverse variance relative weight map is also generated. 

We note that most Q1 papers use MER stacked tiles and catalogues. These tiles are made by MER from VIS images combined with an RMS map, where bad pixels are weighted accordingly. The stacked images produced by VIS and described here, covering the full \ac{FPA}, are not used in the downstream processing.

\subsection{\label{sc:catex} Catalogue extraction}

We extract catalogues using \texttt{SExtractor} \citep{1996A&AS..117..393B} from the stacked images using the \ac{PSF} model and weight map derived above.

\section{\label{sc:AutomatedQC}Automated quality control}

\begin{figure}[ht]
\centering
\includegraphics[width=1.0\hsize]{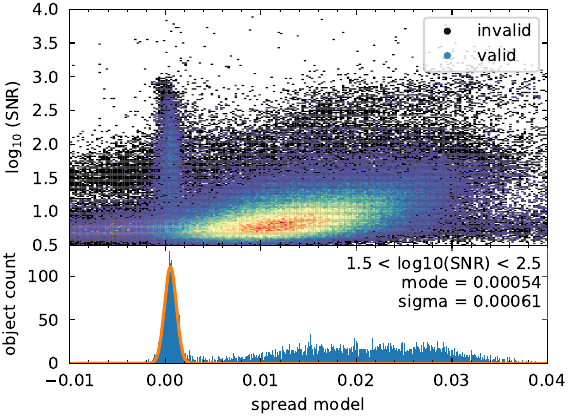}
\caption{\texttt{spread\_model} values for a nominal Q1 exposure. The top panel shows a two-dimensional histogram of \texttt{spread\_model} against S/N (defined as the ratio of \texttt{FLUX\_AUTO} to \texttt{FLUXERR\_AUTO}). The colour scale corresponds to the density of points. The bottom panel shows the histogram of \texttt{spread\_model} values, with the best-fitting Gaussian overlaid in orange.  
\label{fig:Spread-Model-Image}}
\end{figure}

The \ac{VIS PF} is executed on all data, even bad or incomplete images. For each VIS calibrated frame, an \texttt{xml} file is produced containing quality assessment information, in addition to a \texttt{json} file that contains fine-grained quality control information for each of the 144 quadrants. A list of the data-quality control statistics provided is given in \cref{sc:prod-crflags}. They are described in the following sections. 

\subsection{\label{sc:spread-model}\texttt{spread\_model} measurements}

During commissioning, the \texttt{spread-model} parameter was found to be highly correlated with the presence of guiding errors. In \ac{VIS PF}, the  \texttt{FPA\_spreadmodel\_peak} parameter represents the peak of the distribution in \texttt{flux\_auto} versus \texttt{spread\_model} for objects in the range

\begin{equation}
1.5 < \logten\left(\frac{\texttt{FLUX\_AUTO}}{\texttt{FLUXERR\_AUTO}}\right) < 2.5,
\end{equation}
where \texttt{FLUX\_AUTO} and \texttt{FLUXERR\_AUTO} are an estimate of the total flux with the Kron radius \citep{1980ApJS...43..305K}. 

\Cref{fig:Spread-Model-Image} shows \texttt{spread\_model} against S/N (top panel) together with a histogram of \texttt{spread\_model} values with the best-fitting Gaussian overlaid in orange. The locus of unresolved sources at $\texttt{spread\_model}$ at $\sim 0$ is clearly visible. Using \ac{PV} data, and comparing with spacecraft housekeeping and telemetry data, we determined that \texttt{spread\_model} peak values above 0.002 correspond to data affected by guiding errors. The \texttt{FPA\_spreadmodel\_sigma} is the standard deviation of the Gaussian fit described above.

\subsection{\label{sc:score}Image score}

A second image quality metric, the `score' is computed for each exposure using image postage stamps extracted with \texttt{SExtractor}. We compute the median sizes $R^2$ and ellipticities $(e_1,e_2)$ of high-S/N stars\footnote{Stars with peak fluxes at least 10\%\ of the pixel saturation limit are used.} across the FPA, derived from second moments weighted by a \ang{;;0.25} standard deviation Gaussian aperture (see Eqs.~\ref{eq:R2} and \ref{eq:e1e2}. We note that a smaller standard deviation is used to mitigate the impact of \acp{CR}). 

The score expresses how much these values deviate from the typical exposure-to-exposure scatter, as determined from the fine-guidance test campaign during \ac{PV} observations in September 2023. Because guiding errors are larger in the $Y$ direction than in $X$, the $R^2$ and $e_1$ values are negatively correlated. The score is computed as: 
%

\begin{equation}
\text{Score} = \vec{V} \cdot \tens{C}^{-1} \cdot \vec{V}^{\rm T},
\end{equation}
where
\begin{equation}
\vec{V} = \left( R^2 - R^2_{\rm ref},\; 
e_1 - e_{1,\rm ref},\; 
e_2 - e_{2,\rm ref} \right),
\end{equation}
with the reference values
\begin{equation}
R^2_{\rm ref} = 1.923 \,\text{pixels}^{2}, \quad
e_{1,\rm ref} = -0.02313, \quad
e_{2,\rm ref} = 0.00273.
\end{equation}
The covariance matrix is given by
\begin{equation}
\tens{C} = 10^{-6} 
\begin{bmatrix}
542.2  & -118.4 & 0 \\
-118.4 & 44.88  & 0 \\
0      &   0    & 5.332
\end{bmatrix}.
\end{equation}

As the VIS \ac{PSF} is chromatic due to the dichroic beam splitter and the wide VIS bandpass, when observations are made with a population with an atypical colour distribution, the median second moments and therefore the score will be affected. Currently, this colour dependence is not accounted for, as the score mainly serves to identify suspect images. All images with a score greater than 60 undergo a quality control check. Large score values indicate either a proton shower -- which affects the automatic star selection or the second moments -- or a tracking failure.

\subsection{\label{sc:otherqualcontrol}Other quality control statistics}

Finally,  we computed several other quality control statistics. From maps of \acp{CR} detected by \texttt{LaCosmic} (\cref{sc:CR}) we computed \texttt{FPA\_max\_cr} and \texttt{FPA\_fpa\_cr}, which are the percentage of pixels flagged as \acp{CR} in the most affected quadrant and the average percentage of \acp{CR} over the \ac{FPA}, respectively. We also computed the mean astrometric residual over the \ac{FPA}, computed from the difference between the positions of sources in the \ac{FPA} with their counterparts in Gaia DR3 propagated to the epoch of observation.

\section{\label{sc:OnOrbitPerf} Q1 validation}
\subsection{Q1 data quality control}

The Q1 release includes 852 VIS images in four fields: the Euclid Deep Field North (EDF-N), Euclid Deep Field South (EDF-S), Euclid Deep Field Fornax (EDF-F) and single pointing in the Lynds Dark Nebula LDN1641 \citep[``Dark Cloud'', see][]{Q1-TP001}. The Q1 data release contains only data at the depth of the \ac{EWS}, except for the Dark Cloud, which is deeper. 

\begin{figure}[htbp!]
\centering
\includegraphics[width=1.0\hsize]{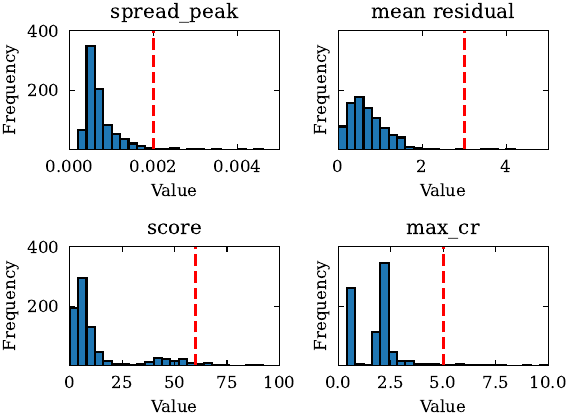}
\caption{Distribution of \texttt{spread\_peak}, \texttt{score}, \texttt{mean\_residual} and \texttt{max\_cr} for all the 852 Q1 VIS images. After the application of these quality control statistics, 836 images remain.}
\label{fig:Q1-DQC.pdf}
\end{figure}

Applying the data-quality control flags described in \cref{sc:AutomatedQC} to the Q1 VIS data, we rejected 16 exposures (1.9\,\%). An exposure is discarded if any of the following conditions are met: (i) one of the three criteria (\texttt{spread\_peak}, \texttt{score}, or \texttt{astrometric residual}) is met, (ii) all quadrants are masked by the \texttt{CR\_REGION} flag, or (iii) the zero-point cannot be computed. \Cref{fig:Q1-DQC.pdf} shows the distribution of Q1 quality control flags, with the dotted line indicating the threshold above which an image is considered potentially unusable. The effect of this image rejection procedure can be seen in the coverage map in the bottom right-hand panel of figure~10 in \citeauthor{Q1-TP001} where shallower depths are a consequence of the rejection of two images affected by guiding errors.

\subsection{\label{sc:MergedCat} Making the merged stellar catalogue}

In this section, we consider the photometric and astrometric characteristics of these data by analysing the 836 VIS calibrated frame catalogues produced by \ac{VIS PF}. These catalogues are extracted from both nominal and short exposure data. Although some of these statistics presented here are computed on an image-by-image basis as part of the normal VIS processing, the aim is to determine the ensemble properties of the entire Q1 dataset and in particular to investigate the characteristics of objects that have been observed multiple times in separate exposures. 

\ac{VIS PF} provides the catalogues as \ac{MEF} files, each with 144 extensions, one for each quadrant. We read each quadrant and compute the position on the VIS focal plane $(x,y)$ in millimetres for each source (this quantity is not present in the Q1 catalogues but will be available in the future), as well as adding a quadrant identifier \texttt{CCD\_QUADID}. The final product is a single catalogue combining all quadrants. Next, we isolate a sample of unsaturated stars in both short exposures and long exposures using the \texttt{FLUX\_RADIUS} -- aperture magnitude (in 13 pixel diameter) relation. We select stars for both exposures using the following criteria: 

\begin{align}
    \text{Selection mask:} \quad & 1.0 < \texttt{FLUX\_RADIUS} < 1.3, \\
    & \texttt{FLAGS} = 0. \notag \\
    \text{Short exposures:} \quad & \quad 16.5 < \texttt{mag} < 21.0, \\
    \text{Long exposures:} \quad & \quad 18.5 < \texttt{mag} < 22.0.
\end{align}

Finally, for each catalogue, these stars are matched to the Gaia DR3 catalogue using a \ang{;;0.2} matching radius. Using the Gaia DR3 proper motions, we compute the necessary correction to proper motion for the matching VIS stars. Next, we positionally merge the matched Gaia-VIS catalogues created above from all observations into a single master catalogue. This catalogue contains all the VIS observations for both short and long exposures for each star. We apply an additional criterion that each group must contain only objects from different \texttt{CCD\_QUADID} and from different exposures. The dither pattern in the \ac{EWS} is large enough to ensure that most object groups contain three or four observations. Unless otherwise noted, all the tests described here use only the long exposures, although we find no significant differences with the short-science exposures.  

\subsection{\label{sc:AP} Astrometric accuracy of the merged Q1 catalogues}
\begin{figure}[htbp!]
\centering
\includegraphics[width=1.0\hsize]{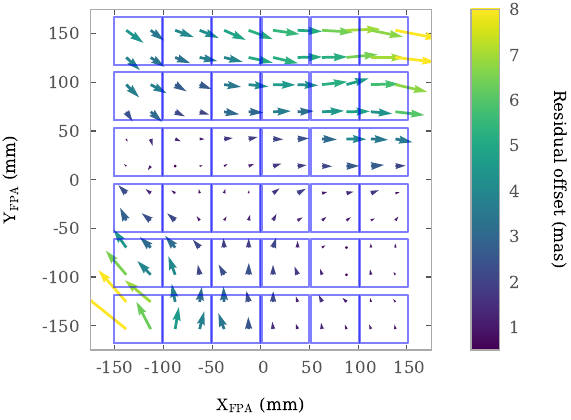}
\caption{Median absolute astrometric precision over the focal plane for validated nominal exposures in Q1. Each arrow denotes the median difference between Gaia stars precessed to the \Euclid observation epoch and the VIS positions. The blue squares indicate the boundaries of each detector.}
\label{fig:astrometry-extern}
\end{figure}

Using the merged catalogue described above, for each group we compute the mean $(x,y)$ position and the mean astrometric coordinates, $(\texttt{ALPHAWIN\_J2000}$,\,$\texttt{DELTAWIN\_J2000})$. Next, these mean positions are binned on the focal plane in a $12\times12$ grid and median residuals with respect to Gaia are computed corresponding to the square root of the sum of the squares of the residuals in RA and Dec. This is shown in \cref{fig:astrometry-extern}. These residuals are smaller than 8\,mas. Over all exposures, the standard deviation of the residuals in RA and Dec is 7\,mas and 8\,mas respectively. We find that the residual field exhibits a non-zero curl. This is because the Q1 distortion matrix has not been updated since the \ac{PV} phase. Since the Q1 processing described here, we have reprocessed larger datasets with a distortion matrix recomputed from the self-cal field at the appropriate epochs, and the circular pattern is no longer present.  

\begin{figure}[htbp!]
\centering
\includegraphics[width=1.0\hsize]{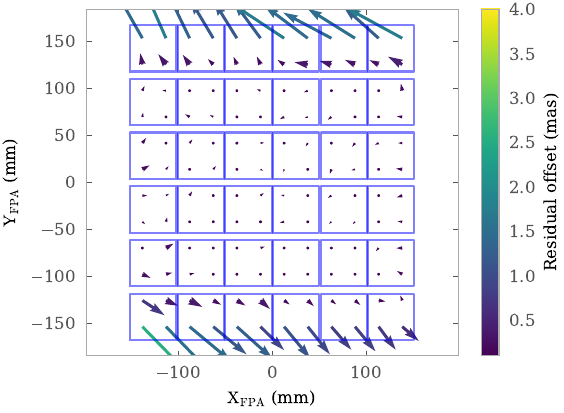}
\caption{Median internal astrometric precision over the focal plane for nominal validated exposures in Q1.}
\label{fig:astrometry-intern}
\end{figure}

We also computed the same statistics internally, this time comparing the positions of the same sources on multiple exposures, shown in \cref{fig:astrometry-intern} (note that the scale of the length of the arrows is not the same between the two plots). We note that the registration over the separate exposures is much better than 1\,mas over most of the field.  

\subsection{\label{sc:photo-scatter} Photometric scatter between observations}

\begin{figure}[htbp!]
\centering
\includegraphics[width=1.0\hsize]{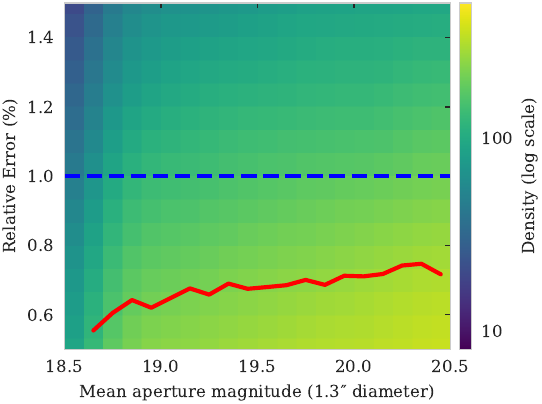}
\caption{Photometric scatter as a function of \mbox{\ang{;;1.3}} diameter aperture magnitudes for validated nominal exposures in Q1. The shaded background shows the number of points as a two-dimensional histogram, with lighter-coloured regions corresponding to more objects.}
\label{fig:magnitude-intern}
\end{figure}

Next, using our merged catalogue, we investigate the photometric scatter between multiple observations of the same object in the nominal exposures over all Q1 data. This is shown in \cref{fig:magnitude-intern}. The dashed blue line  line shows the photometric requirement, and the shaded background the distribution of points as a two-dimensional histogram. For each group of observations, the relative flux error is computed as the group's standard deviation normalised by the group's mean flux and plotted as the solid red line.  Over the magnitude range considered, the relative flux error in percent ranges from 0.6\% to 0.7\%, well within the 1\% requirement. 

As an additional check, we cross-matched the Q1 MER photometric catalogue \citep{Q1-TP004} with this catalogue. We compared the mean 13-pixel diameter aperture flux of each object in each group (corrected to total magnitudes using the \texttt{APERCOR} factor described in \cref{sc:PHO}) with the MER total magnitudes. At $\IE<19.5$ we find that these corrected magnitudes agree with MER fluxes to within  1\%. 

\subsection{\label{sc:counts} VIS galaxy counts}

\begin{figure}[htbp!]
\centering
\includegraphics[width=1.0\hsize]{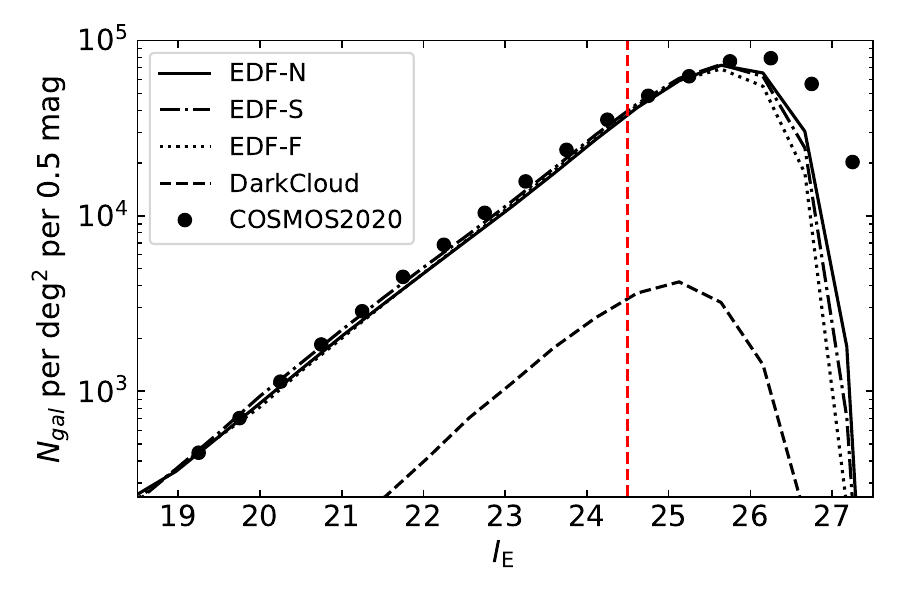}
\caption{Mean magnitude histogram of galaxy counts per square degree from VIS stacks (a single pointing containing four nominal and two short exposures). The mean is given separately for each of the sky patches of the Q1 release, and each is an average of around 40 stacks. The bin width is 0.5.  Galaxies were selected according to a minimum size ($\geq 1.25$ times the \ac{PSF} size) and signal-to-noise ($\geq 10$).  Each VIS stack covers approximately 0.6208\,deg$^2$.  The COSMOS galaxy counts are based on estimated \IE magnitudes, by fitting SEDs over all the COSMOS bands.
\label{fig:magnitude-histogram}}
\end{figure}

Finally, \cref{fig:magnitude-histogram} shows the galaxy counts for the four different sky patches, computed using the \texttt{ mag\_auto} magnitudes extracted from the stacked images. Counts are computed in half-magnitude bins normalised to per square degree. Galaxies were selected from the catalogues by imposing a size ($\geq 1.25$ times the \ac{PSF} size) and signal-to-noise ($\geq 10$) requirement to mimic the object selection for cosmic shear measurements. 

The Dark Cloud region has lower counts because of heavy dust obscuration. In the remaining three Euclid Deep Fields (North, South, and Fornax), the galaxy counts are in excellent agreement, with the turnover magnitudes differing slightly.  The turn-off is at fainter magnitudes where the zodiacal sky background is the lowest, or the furthest away from the ecliptic plane. EDF-N is at the north ecliptic pole, and the ecliptic latitude is lower for EDF-S and EDF-F.

The cumulative galaxy counts at $\IE=24.5$ are around 30\,arcmin$^{-2}$, meeting the general \Euclid requirement.   Our catalogues can be considered complete here, given that the counts turn over at much fainter magnitudes. We also compare with counts computed using a special version of the COSMOS2020 catalogue \citep{2022ApJS..258...11W} where the \IE flux of each source has been computed from the best-fitting \ac{SED} following the methods outlined in \cite{2020MNRAS.494..199S}. The agreement is excellent, especially given that our area computation does not account for area lost by bright stars or image artefacts. 


We note that in the stacked images (whether by \ac{VIS PF} or by the downstream \ac{MER PF}), the survey depth varies across the image, with the depth variation dominated by the number of nominal exposures.  The bulk of the wide survey consists of pointings with three or four nominal exposures. Analysis of the EDF-N field catalogues shows that the turnover point for the three- and four-image stack depths differs by $\sim$ 0.2\,mag, and the EDF-N counts in \cref{fig:magnitude-histogram} lies between the three- and four-image stack curves.  This same behaviour has been observed in shallower ecliptic latitudes with more zodiacal light than in EDF-N.

\section{\label{sc:caveats} Known caveats}

As part of the Q1 validation process, we compared the fluxes of identical sources in both short and nominal exposures. Fluxes were normalised to the exposure times reported by the \texttt{EXPTIME} header keyword, and then compared between short and long exposures. This test indicates that the fluxes in the normalised long exposures are approximately \SI{2}{\percent} lower than in normalised short exposures. As described in \cref{sc:ILL}, the exposure time across the \ac{FPA} is non-uniform because of the shutter function.  This non-uniformity is corrected for by the respective short-exposure  and nominal-exposure large-scale flats constructed from relative stellar photometry. However, in the Q1 dataset described here, the absolute normalisation of the large-scale flat is the average exposure time of the \ac{FPA}, which does not consider the relative flux differences induced by the shutter. For future releases, we will first scale the fluxes between the short and nominal exposures to remove this discrepancy. Second, once an accurate shutter-movement model is available, we will remove the shutter-dependent exposure time variation from the short and nominal exposures, and the flux differences will disappear. 

We also compared the MER photometric catalogue \citep{Q1-TP004} with the calibrated frames presented here. We compared the mean flux of each object, measured within a 13-pixel diameter aperture, of each object in each group (corrected to total magnitudes using the \texttt{APERCOR} factor as described in \cref{sc:PHO}) with the MER total magnitudes. For sources with $\IE<19.5$ we find that these corrected magnitudes agree with MER fluxes to within  \SI{1}{\percent}. The origin of this discrepancy is under investigation, but this test provides a useful upper limit on the magnitude of the effect described above on the final photometric measurements.

\section{\label{sc:Conc} Conclusions}
This paper presents the automatic pipeline that has been developed to process data from \Euclid's VIS camera, \ac{VIS PF}. This software computes master calibration data for instrumental effects present in VIS data and uses these calibration products to produce science-ready data from raw VIS images. The \ac{VIS PF} also provides a detailed set of quality control parameters and flags for each image that allows robust data selection to be made at both the image and pixel level. We demonstrate the performance of \ac{VIS PF} on data taken during the \ac{PV} phase in August 2023 and show that the principal mission requirements are met. We described the VIS Q1 data release, and showed the excellent quality of the processed and validated Q1 data. Images have a highly stable \ang{;;0.16} \ac{FWHM} and the image-to-image photometric scatter is less than \SI{1}{\percent}. The processed VIS images (and the associated science papers in this Q1 data release) provide a glimpse of the revolutionary science that these wide-field high-resolution images will achieve. 

\begin{acknowledgements}
This work was made possible by utilising the CANDIDE cluster at the Institut d'Astrophysique de Paris. The cluster was funded through grants from the PNCG, CNES, DIM-ACAV, the Euclid Consortium, and the Danish National Research Foundation Cosmic Dawn Center (DNRF140). It is maintained by Stephane Rouberol. We thank the referee for their comments which improved our paper. 
\AckEC  
\AckQone

\end{acknowledgements}

\bibliography{Euclid,manual-refs}

\begin{appendix}

\section{\label{sc:image-notation}{Ellipticity and size measurement}}

To avoid ambiguities, this section presents the ellipticity and size notation used in this paper, following the standard definitions adopted in the \Euclid requirement documents.

 To quantify the PSF size, we rely on the squared radius $R^2$, as defined in equation~(1) of \cite{masseyOriginsWeakLensing2013}.  We use a circular Gaussian weight of width $\sigma=\ang{;;0.75}$  for an object at coordinate $\vec{x}$:

\begin{equation}
w(\vec{x})
= \frac{1}{2\pi\sigma^2}
  \exp\brackets{
    -\tfrac{(x_1 - x_1^{\rm cen})^2 + (x_2 - x_2^{\rm cen})^2}{2\sigma^2}
  }. \label{eq:weight}
\end{equation}

The weighted flux is then
\begin{equation}
F^{(0)}
= \int w(\vec{x})\,f(\vec{x})\,\mathrm{d}^2\vec{x}. \label{eq:flux}
\end{equation}

The centroid is re‐evaluated four times, starting from the vignette centre:

\begin{equation}
x_{i}^{\rm cen}
= \frac{1}{F^{(0)}}
  \int w(\vec{x})\,x_{i}\,f(\vec{x})\,\mathrm{d}^2\vec{x}
  \quad(i=1,2). \label{eq:centroid}
\end{equation}

Using the same weight, the second‐order (quadrupole) moments are
\begin{equation}
Q_{ij}
= \frac{1}{F^{(0)}}
  \int w(\vec{x})\,(x_{i}-x_{i}^{\rm cen})(x_{j}-x_{j}^{\rm cen})\,f(\vec{x})\,\mathrm{d}^2\vec{x}. \label{eq:Qij}
\end{equation}

The total size is expressed through the squared radius
\begin{equation}
R^{2} = Q_{11} + Q_{22}. \label{eq:R2}
\end{equation}

The complex ellipticity components are
\begin{subequations}\label{eq:ellip}
\begin{align}
e_{1} &= \frac{Q_{11}-Q_{22}}{Q_{11}+Q_{22}},\\
e_{2} &= \frac{2\,Q_{12}}{Q_{11}+Q_{22}}.
\end{align}
\label{eq:e1e2}
\end{subequations}

Their magnitude gives the total ellipticity,
\begin{equation}
e = \sqrt{e_{1}^{2} + e_{2}^{2}}. \label{eq:ellip_tot}
\end{equation}

\section{\label{sc:q1-products}{A brief description of VIS Q1 data products}}

Q1 products are distributed through the \fnurl{Euclid Science Archive (SAS).}{https://easidr.esac.esa.int/sas} VIS Q1 products are fully described in the \fnurl{data product description documentation.}{https://euclid.esac.esa.int/dr/q1/dpdd/visdpd/visindex.html} The SAS includes both processed and raw VIS images from nominal and short exposures. Catalogues extracted from the calibrated frames are also provided. Both catalogues and images are delivered in the FITS \citep{1981A&AS...44..363W,2010A&A...524A..42P} format.  

\subsection{\label{sc:q1-calframe-desc}{The VIS calibrated quad frame}}

In the \fnurl{VIS calibrated quad frame}{https://euclid.esac.esa.int/dr/q1/dpdd/visdpd/dpcards/vis_calibratedquadframe.html}\textsuperscript{\kern-0.2em,\kern-0.0em}\footnote{We note that here we refer to the `VIS calibrated quad frame'; the `VIS calibrated frame' is an older data product where the different detector quadrants are grouped into \acp{CCD}.} data product, three kinds of FITS files are supplied. In each file, the quadrant arrangement follows \cref{fig:focal-plane-layout}. The \texttt{DET} (for `detrended') \ac{MEF} file contains three extensions \texttt{SCI}, \texttt{RMS} and \texttt{FLG} for each of the 144 quadrants. The \texttt{SCI} extension corresponds to the VIS science image. Both nominal and short exposures are supplied, in \acp{ADU} with a gain of 3.48\,e$^{-}$ for each quadrant and a zero-point of 24.57\,s$^{-1}$.  

The \texttt{RMS} extension is a noise map constructed by summing in quadrature the science image photon noise and quadrant readout noise (both in electrons), then converting the total to 
\acp{ADU}.\footnote{The readout noise is computed separately from the median of several thousand measurements of the standard deviation of the quadrant serial overscan.}

The \texttt{FLG}extension is a quality-control image delivered as a signed integer. The \texttt{INVALID} is a convenience flag that contains most of the bad regions listed below. The code snippet below shows how to read this file and select only object pixels suitable for science (\texttt{good\_objects}). 

\begin{lstlisting}[caption={Example code to select object pixels unaffected by invalid pixels (including a full flag list).}, label={lst:good_objects_selection}]

with fits.open(fits_file) as hdul:
    flag_array = hdul[3].data.astype(np.int32)

# Define bad flags with their meanings
bad_flag_definitions = {
    0: "INVALID", 1: "HOT", 2: "COLD", 
    3: "SAT", 4: "COSMIC", 5: "GHOST", 
    7: "BAD_COLUMN", 8: "BAD_CLUSTER",
    9: "CR_REGION", 12: "OVRCOL", 
    15: "CHARINJ", 17: "SATXTALKGHOST",
    21: "ADCMAX", 22: "NO_DATA"
}

bad_flags = sum(1 << i for i in bad_flag_definitions)  # Combine all bad flags
invalid_flag = 1 << 0  # INVALID flag

object_flags = (1 << 18) | (1 << 24)  # STARSIGNAL, OBJECTS

# Identify pixels that are objects
is_object = (flag_array & object_flags) != 0  # True for OBJECTS or STARSIGNAL pixels

# Identify pixels that are not marked as INVALID

is_valid = (flag_array & invalid_flag) == 0  
good_objects = np.logical_and(is_object, is_valid)

\end{lstlisting}

Finally, two additional FITS files are supplied: the weight map and background map (indicated by \texttt{WGT} and \texttt{BKG} in the filename). The weight map is simply the small-scale flat (Sect.~\ref{sc:flat}) where all pixels flagged as \texttt{INVALID} are set to zero. The background file is the \texttt{NoiseChisel} (Sect.~\ref{sc:bkg}) background map. 

\subsection{\label{sc:q1-catalogue-desc}{The VIS calibrated quad frame catalogue}}

The VIS calibrated quad frame catalogue is a \texttt{FITS} table catalogue with 144 extensions produced by \texttt{SExtractor}. The \ac{DPDD} provides a complete list of parameters that are almost all standard \texttt{SExtractor} parameters, except for calibrated flux measurements in microjanskys, which are indicated by the suffix \texttt{\_CAL}. 

\section{\label{sc:prod-crflags} Image quality control statistics}

Each VIS calibrated frame is supplied with an \texttt{xml} file that contains the following quantities:  

\begin{itemize}
    \item \textbf{\texttt{FPA\_spreadmodel\_peak}} -- the peak of the \texttt{spread\_model} distribution (\cref{sc:spread-model});
    \item \textbf{\texttt{FPA\_spreadmodel\_sigma}} -- the width of the \texttt{spread\_model} distribution (\cref{sc:spread-model}); 
    \item \textbf{\texttt{FPA\_max\_cr}} -- percentage of pixels impacted by \acp{CR} in the most affected quadrant (\cref{sc:otherqualcontrol});
    \item \textbf{\texttt{FPA\_fpa\_cr}} -- average percentage of pixels flagged as \acp{CR}  over the \ac{FPA} (\cref{sc:otherqualcontrol});
    \item \textbf{\texttt{FPA\_cr\_regions}} -- the number of quadrants masked by the \texttt{CR\_REGION} flag  (the number of quadrants where more than 5\% of pixels are flagged as \acp{CR});
    \item \textbf{\texttt{FPA\_mean\_residual}} -- The mean astrometric residual averaged over the \ac{FPA}; 
    \item \textbf{\texttt{FPA\_e1\_med}} -- the median value of $e_1$ of selected stars (\cref{sc:score});
    \item \textbf{\texttt{FPA\_e2\_med}} -- the median value $e_2$ of selected stars (\cref{sc:score});
    \item \textbf{\texttt{FPA\_r2\_med}} -- the median value of the $R^2$ of selected stars (\cref{sc:score});
    \item \textbf{\texttt{SCORE}} -- the overall image quality score (\cref{sc:score}). 
\end{itemize}

\section{\label{sc:q1-configuration-files}{Configuration files}}

For the \texttt{SExtractor} configuration, most parameters were kept close to their defaults. Aperture photometry is measured in four apertures of 7, 13, 26 and 50 pixels in diameter. \texttt{DETECT\_THRESH} was set to 1.5. For \texttt{PSFEx}, we chose \texttt{BASIS\_TYPE=PIXEL\_AUTO} and \texttt{PSF\_SIZE=21,21} and fitted the \ac{PSF}  variation with a second-degree polynomial with \texttt{PSFVAR\_DEGREES=2}.

\section{\label{sc:VIS_throughput}{VIS throughput}}
\begin{figure}[htbp!]
\centering
\includegraphics[width=1.0\hsize]{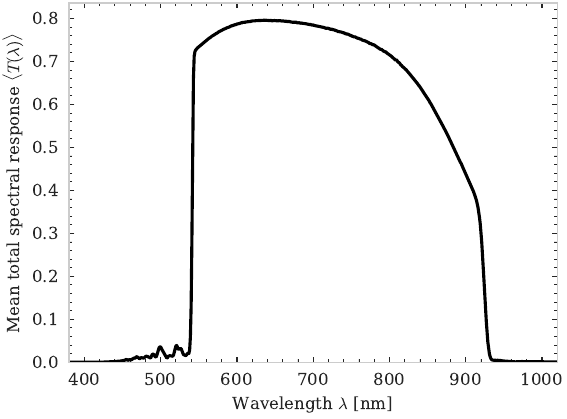}
\caption{VIS throughput, showing  the mean photon-to-electron spectral response of the \textit{Euclid} VIS channel, sampled at 1\,nm intervals over the range $380$–$1020$\,nm.}
\label{fig:VIS_response}
\end{figure}
For reference, Fig.~\ref{fig:VIS_response} shows the mean photon-to-electron spectral response of the \textit{Euclid} VIS channel, sampled at 1\,nm intervals over the range $380$–$1020$\,nm.  The response is defined as

\begin{equation}
\ave{T(\lambda)} = R_{\mathrm{Ag}}^{4}(\lambda)\, 
                   R_{\mathrm{hyb}}^{2}(\lambda)\, 
                   R_{\mathrm{dich}}(\lambda)\, 
                   \ave{\mathrm{QE}}(\lambda),
\end{equation}

\noindent where $R_{\mathrm{Ag}}(\lambda)$ and $R_{\mathrm{hyb}}(\lambda)$ are the reflectances of the four silver-coated telescope mirrors and the two hybrid-coated VIS fold mirrors, respectively.  
The dichroic reflectance is obtained as $R_{\mathrm{dich}}(\lambda)=1-T_{\mathrm{dich}}(\lambda)$ from its tabulated transmission, assuming negligible absorption.  
The coating and dichroic data are taken from ground testing.  
The mean CCD quantum efficiency $\langle\mathrm{QE}\rangle(\lambda)$ read from the \Euclid mission database is averaged over all detectors and interpolated with a cubic spline to the same 1\,nm grid.  The result quantifies the wavelength-dependent throughput of the complete VIS optical chain upstream of the detector.

\end{appendix}

\end{document}

%% file: acronym.tex
\acrodef{AA}{azimuth or $\alpha$ angle}
\acrodef{ADC}{analogue-to-digital converter}
\acrodef{ADU}{analogue-to-digital unit}
\acrodef{AOI}{angle of incidence}
\acrodef{ASIC}{application specific integrated circuit}
\acrodef{BFE}{brighter-fatter effect}
\acrodef{CaLA}{camera-lens assembly}
\acrodef{CCD}{charge-coupled device}
\acrodef{CoLA}{corrector-lens assembly}
\acrodef{CDS}{Correlated Double Sampling}
\acrodef{CFC}{cryo-flex cable}
\acrodef{CFHT}{Canada-France-Hawaii Telescope}
\acrodef{CFRP}{carbon-fibre reinforced plastic}
\acrodef{CGH}{computer-generated hologram}
\acrodef{CME}{coronal mass ejection}
\acrodef{CNES}{Centre National d'Etude Spacial}
\acrodef{CPPM}{Centre de Physique des Particules de Marseille}
\acrodef{CPU}{central processing unit}
\acrodef{CR}{cosmic ray}
\acrodef{CTE}{coefficient of thermal expansion}
\acrodef{CME}{coronal mass ejection}
\acrodef{CTI}{charge-transfer inefficiency}
\acrodef{DCU}{Detector Control Unit}
\acrodef{DES}{Dark Energy Survey}
\acrodef{DPU}{Data Processing Unit}
\acrodef{DS}{Detector System}
\acrodef{EDS}{Euclid Deep Survey}
\acrodef{EE}{encircled energy}
\acrodef{EPER}{extended pixel-edge response}
\acrodef{ESA}{European Space Agency}
\acrodef{ESP}{Emission of Solar Protons}
\acrodef{ECSS}{European Cooperation for Space Standardization}
\acrodef{EWS}{Euclid Wide Survey}
\acrodef{FDIR}{Fault Detection, Isolation and Recovery}
\acrodef{FGS}{fine guidance sensor}
\acrodef{FOM}{figure of merit}
\acrodef{FOV}{field of view}
\acrodef{FPA}{focal-plane array}
\acrodef{FPR}{false positive rate}
\acrodef{FWA}{filter-wheel assembly}
\acrodef{FWC}{full-well capacity}
\acrodef{FWHM}{full width at half maximum}
\acrodef{GOES}{Geostationary Operational Environmental Satellites}
\acrodef{GCR}{Galactic cosmic ray}
\acrodef{GWA}{grism-wheel assembly}
\acrodef{H2RG}{HAWAII-2RG}
\acrodef{HST}{\textit{Hubble} Space Telescope}
\acrodef{IP2I}{Institut de Physique des 2 Infinis de Lyon}
\acrodef{JWST}{{\em James Webb} Space Telescope}
\acrodef{IAD}{ion-assisted deposition}
\acrodef{ICU}{Instrument Control Unit}
\acrodef{IPC}{inter-pixel capacitance}
\acrodef{ISES}{International Space Environmental Services}
\acrodef{JWST}{\textit{James Webb} Space Telescope}
\acrodef{LAM}{Laboratoire d'Astrophysique de Marseille}
\acrodef{LED}{light-emitting diode}
\acrodef{LSB}{low surface brightness}
\acrodef{MACC}{multiple accumulated}
\acrodef{MEF}{multi-extension FITS}
\acrodef{MER PF}{MER processing function}
\acrodef{MLI}{multi-layer insulation}
\acrodef{MMU}{Mass Memory Unit}
\acrodef{MPE}{Max-Planck-Institut für extraterrestrische Physik}
\acrodef{MPIA}{Max-Planck-Institut für Astronomie}
\acrodef{NA}{numerical aperture}
\acrodef{NASA}{National Aeronautic and Space Administration}
\acrodef{NIEL}{non-ionising energy loss}
\acrodef{JPL}{NASA Jet Propulsion Laboratory}
\acrodef{MZ-CGH}{multi-zonal computer-generated hologram}
\acrodef{NI-CU}{NISP calibration unit}
\acrodef{NI-OA}{near-infrared optical assembly}
\acrodef{NI-GWA}{NISP Grism Wheel Assembly}
\acrodef{NIR}{near-infrared}
\acrodef{NISP}{Near-Infrared Spectrometer and Photometer}
\acrodef{NOAA}{National Oceanic and Atmospheric Administration}
\acrodef{PA}{position angle}
\acrodef{PARMS}{plasma-assisted reactive magnetron sputtering}
\acrodef{PLM}{payload module}
\acrodef{PRNU}{pixel-response non-uniformity}
\acrodef{PTC}{photon transfer curve}
\acrodef{PTFE}{polytetrafluoroethylene}
\acrodef{PV}{performance verification}
\acrodef{PWM}{pulse-width modulation}
\acrodef{PSF}{point spread function}
\acrodef{QE}{quantum efficiency}
\acrodef{QF}{quality factor}
\acrodef{ROE}{readout electronic block unit}
\acrodef{ROI}{region of interest}
\acrodef{ROIC}{readout-integrated circuit}
\acrodef{ROS}{reference observing sequence}
\acrodef{SAA}{Solar aspect angle}
\acrodef{SCA}{sensor chip array}
\acrodef{SCE}{sensor chip electronic}
\acrodef{SCS}{sensor chip system}
\acrodef{SGS}{science ground segment}
\acrodef{SGPS}{Solar and Galactic Proton Sensor}
\acrodef{SHS}{Shack-Hartmann sensor}
\acrodef{SNR}[S/N]{signal-to-noise ratio}
\acrodef{SED}{spectral energy distribution}
\acrodef{SiC}{silicon carbide}
\acrodef{SEP}{Solar energetic particle}
\acrodef{SSN}{Sunspot number}
\acrodef{STIX}{Spectrometer/Telescope for Imaging X-rays}
\acrodef{SolO}{Solar Orbiter}
\acrodef{TP}{trap pumping}
\acrodef{TPR}{true positive rate}
\acrodef{UTR}{up-the-ramp}
\acrodef{SVM}{service module}
\acrodef{VIS}{visible imager}
\acrodef{VIS PF}{VIS processing function}
\acrodef{WD}{white dwarf}
\acrodef{WCS}{world coordinate system}
\acrodef{WFE}{wavefront error}
\acrodef{ZP}{zero point}
\acrodef{DPDD}{Data Product Description Document}

%% file: authors.tex
\newcommand{\orcid}[1]{} 
\author{Euclid Collaboration: H.~J.~McCracken\orcid{0000-0002-9489-7765}\thanks{\email{hjmcc@iap.fr}}\inst{\ref{aff1}}
\and K.~Benson\inst{\ref{aff2}}
\and C.~Dolding\orcid{0009-0003-7199-6108}\inst{\ref{aff2}}
\and T.~Flanet\inst{\ref{aff1}}
\and C.~Grenet\inst{\ref{aff1}}
\and O.~Herent\inst{\ref{aff1}}
\and P.~Hudelot\inst{\ref{aff1}}
\and C.~Laigle\orcid{0009-0008-5926-818X}\inst{\ref{aff1}}
\and G.~Leroy\orcid{0009-0004-2523-4425}\inst{\ref{aff3},\ref{aff4}}
\and P.~Liebing\inst{\ref{aff2}}
\and R.~Massey\orcid{0000-0002-6085-3780}\inst{\ref{aff4}}
\and S.~Mottet\inst{\ref{aff1}}
\and R.~Nakajima\orcid{0009-0009-1213-7040}\inst{\ref{aff5}}
\and H.~N.~Nguyen-Kim\inst{\ref{aff1}}
\and J.~W.~Nightingale\orcid{0000-0002-8987-7401}\inst{\ref{aff6}}
\and J.~Skottfelt\orcid{0000-0003-1310-8283}\inst{\ref{aff7}}
\and L.~C.~Smith\orcid{0000-0002-3259-2771}\inst{\ref{aff8}}
\and F.~Soldano\inst{\ref{aff1}}
\and E.~Vilenius\orcid{0000-0002-6184-7681}\inst{\ref{aff2}}
\and M.~Wander\inst{\ref{aff7}}
\and M.~von~Wietersheim-Kramsta\orcid{0000-0003-4986-5091}\inst{\ref{aff4},\ref{aff3}}
\and M.~Akhlaghi\orcid{0000-0003-1710-6613}\inst{\ref{aff9}}
\and H.~Aussel\orcid{0000-0002-1371-5705}\inst{\ref{aff10}}
\and S.~Awan\inst{\ref{aff2}}
\and R.~Azzollini\orcid{0000-0002-0438-0886}\inst{\ref{aff2}}
\and A.~Basset\inst{\ref{aff11}}
\and G.~P.~Candini\orcid{0000-0001-9481-8206}\inst{\ref{aff2}}
\and P.~Casenove\orcid{0009-0006-6736-1670}\inst{\ref{aff11}}
\and M.~Cropper\orcid{0000-0003-4571-9468}\inst{\ref{aff2}}
\and H.~Hoekstra\orcid{0000-0002-0641-3231}\inst{\ref{aff12}}
\and H.~Israel\orcid{0000-0002-3045-4412}\inst{\ref{aff13}}
\and A.~Khalil\inst{\ref{aff2}}
\and K.~Kuijken\orcid{0000-0002-3827-0175}\inst{\ref{aff12}}
\and Y.~Mellier\inst{\ref{aff14},\ref{aff1}}
\and L.~Miller\orcid{0000-0002-3376-6200}\inst{\ref{aff15}}
\and S.-M.~Niemi\inst{\ref{aff16}}
\and M.~J.~Page\orcid{0000-0002-6689-6271}\inst{\ref{aff2}}
\and K.~Paterson\orcid{0000-0001-8340-3486}\inst{\ref{aff17}}
\and M.~Schirmer\orcid{0000-0003-2568-9994}\inst{\ref{aff17}}
\and N.~A.~Walton\orcid{0000-0003-3983-8778}\inst{\ref{aff8}}
\and A.~Zacchei\orcid{0000-0003-0396-1192}\inst{\ref{aff18},\ref{aff19}}
\and J.~P.~L.~G.~Barrios\orcid{0000-0002-5605-0029}\inst{\ref{aff4},\ref{aff20}}
\and T.~Erben\inst{\ref{aff5}}
\and R.~Hayes\inst{\ref{aff4}}
\and J.~A.~Kegerreis\orcid{0000-0001-5383-236X}\inst{\ref{aff21},\ref{aff4}}
\and D.~J.~Lagattuta\orcid{0000-0002-7633-2883}\inst{\ref{aff3},\ref{aff4}}
\and A.~Lan\c{c}on\orcid{0000-0002-7214-8296}\inst{\ref{aff22}}
\and N.~Aghanim\orcid{0000-0002-6688-8992}\inst{\ref{aff23}}
\and B.~Altieri\orcid{0000-0003-3936-0284}\inst{\ref{aff24}}
\and A.~Amara\inst{\ref{aff25}}
\and S.~Andreon\orcid{0000-0002-2041-8784}\inst{\ref{aff26}}
\and P.~N.~Appleton\orcid{0000-0002-7607-8766}\inst{\ref{aff27},\ref{aff28}}
\and N.~Auricchio\orcid{0000-0003-4444-8651}\inst{\ref{aff29}}
\and C.~Baccigalupi\orcid{0000-0002-8211-1630}\inst{\ref{aff19},\ref{aff18},\ref{aff30},\ref{aff31}}
\and M.~Baldi\orcid{0000-0003-4145-1943}\inst{\ref{aff32},\ref{aff29},\ref{aff33}}
\and A.~Balestra\orcid{0000-0002-6967-261X}\inst{\ref{aff34}}
\and S.~Bardelli\orcid{0000-0002-8900-0298}\inst{\ref{aff29}}
\and P.~Battaglia\orcid{0000-0002-7337-5909}\inst{\ref{aff29}}
\and A.~N.~Belikov\inst{\ref{aff35},\ref{aff36}}
\and R.~Bender\orcid{0000-0001-7179-0626}\inst{\ref{aff37},\ref{aff38}}
\and F.~Bernardeau\inst{\ref{aff39},\ref{aff1}}
\and A.~Biviano\orcid{0000-0002-0857-0732}\inst{\ref{aff18},\ref{aff19}}
\and A.~Bonchi\orcid{0000-0002-2667-5482}\inst{\ref{aff40}}
\and E.~Branchini\orcid{0000-0002-0808-6908}\inst{\ref{aff41},\ref{aff42},\ref{aff26}}
\and M.~Brescia\orcid{0000-0001-9506-5680}\inst{\ref{aff43},\ref{aff44}}
\and J.~Brinchmann\orcid{0000-0003-4359-8797}\inst{\ref{aff45},\ref{aff46}}
\and S.~Camera\orcid{0000-0003-3399-3574}\inst{\ref{aff47},\ref{aff48},\ref{aff49}}
\and G.~Ca\~nas-Herrera\orcid{0000-0003-2796-2149}\inst{\ref{aff16},\ref{aff50},\ref{aff12}}
\and V.~Capobianco\orcid{0000-0002-3309-7692}\inst{\ref{aff49}}
\and C.~Carbone\orcid{0000-0003-0125-3563}\inst{\ref{aff51}}
\and J.~Carretero\orcid{0000-0002-3130-0204}\inst{\ref{aff52},\ref{aff53}}
\and S.~Casas\orcid{0000-0002-4751-5138}\inst{\ref{aff54}}
\and F.~J.~Castander\orcid{0000-0001-7316-4573}\inst{\ref{aff55},\ref{aff56}}
\and M.~Castellano\orcid{0000-0001-9875-8263}\inst{\ref{aff57}}
\and G.~Castignani\orcid{0000-0001-6831-0687}\inst{\ref{aff29}}
\and S.~Cavuoti\orcid{0000-0002-3787-4196}\inst{\ref{aff44},\ref{aff58}}
\and K.~C.~Chambers\orcid{0000-0001-6965-7789}\inst{\ref{aff59}}
\and A.~Cimatti\inst{\ref{aff60}}
\and C.~Colodro-Conde\inst{\ref{aff61}}
\and G.~Congedo\orcid{0000-0003-2508-0046}\inst{\ref{aff62}}
\and C.~J.~Conselice\orcid{0000-0003-1949-7638}\inst{\ref{aff63}}
\and L.~Conversi\orcid{0000-0002-6710-8476}\inst{\ref{aff64},\ref{aff24}}
\and Y.~Copin\orcid{0000-0002-5317-7518}\inst{\ref{aff65}}
\and F.~Courbin\orcid{0000-0003-0758-6510}\inst{\ref{aff66},\ref{aff67}}
\and H.~M.~Courtois\orcid{0000-0003-0509-1776}\inst{\ref{aff68}}
\and A.~Da~Silva\orcid{0000-0002-6385-1609}\inst{\ref{aff69},\ref{aff70}}
\and R.~da~Silva\orcid{0000-0003-4788-677X}\inst{\ref{aff57},\ref{aff40}}
\and H.~Degaudenzi\orcid{0000-0002-5887-6799}\inst{\ref{aff71}}
\and G.~De~Lucia\orcid{0000-0002-6220-9104}\inst{\ref{aff18}}
\and A.~M.~Di~Giorgio\orcid{0000-0002-4767-2360}\inst{\ref{aff72}}
\and J.~Dinis\orcid{0000-0001-5075-1601}\inst{\ref{aff69},\ref{aff70}}
\and H.~Dole\orcid{0000-0002-9767-3839}\inst{\ref{aff23}}
\and F.~Dubath\orcid{0000-0002-6533-2810}\inst{\ref{aff71}}
\and X.~Dupac\inst{\ref{aff24}}
\and S.~Dusini\orcid{0000-0002-1128-0664}\inst{\ref{aff73}}
\and A.~Ealet\orcid{0000-0003-3070-014X}\inst{\ref{aff65}}
\and S.~Escoffier\orcid{0000-0002-2847-7498}\inst{\ref{aff74}}
\and M.~Fabricius\orcid{0000-0002-7025-6058}\inst{\ref{aff37},\ref{aff38}}
\and M.~Farina\orcid{0000-0002-3089-7846}\inst{\ref{aff72}}
\and R.~Farinelli\inst{\ref{aff29}}
\and S.~Ferriol\inst{\ref{aff65}}
\and F.~Finelli\orcid{0000-0002-6694-3269}\inst{\ref{aff29},\ref{aff75}}
\and P.~Fosalba\orcid{0000-0002-1510-5214}\inst{\ref{aff56},\ref{aff55}}
\and S.~Fotopoulou\orcid{0000-0002-9686-254X}\inst{\ref{aff76}}
\and N.~Fourmanoit\orcid{0009-0005-6816-6925}\inst{\ref{aff74}}
\and M.~Frailis\orcid{0000-0002-7400-2135}\inst{\ref{aff18}}
\and E.~Franceschi\orcid{0000-0002-0585-6591}\inst{\ref{aff29}}
\and S.~Galeotta\orcid{0000-0002-3748-5115}\inst{\ref{aff18}}
\and K.~George\orcid{0000-0002-1734-8455}\inst{\ref{aff38}}
\and W.~Gillard\orcid{0000-0003-4744-9748}\inst{\ref{aff74}}
\and B.~Gillis\orcid{0000-0002-4478-1270}\inst{\ref{aff62}}
\and C.~Giocoli\orcid{0000-0002-9590-7961}\inst{\ref{aff29},\ref{aff33}}
\and P.~G\'omez-Alvarez\orcid{0000-0002-8594-5358}\inst{\ref{aff77},\ref{aff24}}
\and J.~Gracia-Carpio\inst{\ref{aff37}}
\and B.~R.~Granett\orcid{0000-0003-2694-9284}\inst{\ref{aff26}}
\and A.~Grazian\orcid{0000-0002-5688-0663}\inst{\ref{aff34}}
\and F.~Grupp\inst{\ref{aff37},\ref{aff38}}
\and L.~Guzzo\orcid{0000-0001-8264-5192}\inst{\ref{aff78},\ref{aff26},\ref{aff79}}
\and M.~Hailey\inst{\ref{aff2}}
\and S.~V.~H.~Haugan\orcid{0000-0001-9648-7260}\inst{\ref{aff80}}
\and J.~Hoar\inst{\ref{aff24}}
\and W.~Holmes\inst{\ref{aff81}}
\and F.~Hormuth\inst{\ref{aff82}}
\and A.~Hornstrup\orcid{0000-0002-3363-0936}\inst{\ref{aff83},\ref{aff84}}
\and K.~Jahnke\orcid{0000-0003-3804-2137}\inst{\ref{aff17}}
\and M.~Jhabvala\inst{\ref{aff85}}
\and B.~Joachimi\orcid{0000-0001-7494-1303}\inst{\ref{aff86}}
\and E.~Keih\"anen\orcid{0000-0003-1804-7715}\inst{\ref{aff87}}
\and S.~Kermiche\orcid{0000-0002-0302-5735}\inst{\ref{aff74}}
\and A.~Kiessling\orcid{0000-0002-2590-1273}\inst{\ref{aff81}}
\and M.~Kilbinger\orcid{0000-0001-9513-7138}\inst{\ref{aff10}}
\and B.~Kubik\orcid{0009-0006-5823-4880}\inst{\ref{aff65}}
\and M.~K\"ummel\orcid{0000-0003-2791-2117}\inst{\ref{aff38}}
\and M.~Kunz\orcid{0000-0002-3052-7394}\inst{\ref{aff88}}
\and H.~Kurki-Suonio\orcid{0000-0002-4618-3063}\inst{\ref{aff89},\ref{aff90}}
\and Q.~Le~Boulc'h\inst{\ref{aff91}}
\and A.~M.~C.~Le~Brun\orcid{0000-0002-0936-4594}\inst{\ref{aff92}}
\and D.~Le~Mignant\orcid{0000-0002-5339-5515}\inst{\ref{aff93}}
\and S.~Ligori\orcid{0000-0003-4172-4606}\inst{\ref{aff49}}
\and P.~B.~Lilje\orcid{0000-0003-4324-7794}\inst{\ref{aff80}}
\and V.~Lindholm\orcid{0000-0003-2317-5471}\inst{\ref{aff89},\ref{aff90}}
\and I.~Lloro\orcid{0000-0001-5966-1434}\inst{\ref{aff94}}
\and G.~Mainetti\orcid{0000-0003-2384-2377}\inst{\ref{aff91}}
\and D.~Maino\inst{\ref{aff78},\ref{aff51},\ref{aff79}}
\and E.~Maiorano\orcid{0000-0003-2593-4355}\inst{\ref{aff29}}
\and O.~Mansutti\orcid{0000-0001-5758-4658}\inst{\ref{aff18}}
\and S.~Marcin\inst{\ref{aff95}}
\and O.~Marggraf\orcid{0000-0001-7242-3852}\inst{\ref{aff5}}
\and M.~Martinelli\orcid{0000-0002-6943-7732}\inst{\ref{aff57},\ref{aff96}}
\and N.~Martinet\orcid{0000-0003-2786-7790}\inst{\ref{aff93}}
\and F.~Marulli\orcid{0000-0002-8850-0303}\inst{\ref{aff97},\ref{aff29},\ref{aff33}}
\and D.~C.~Masters\orcid{0000-0001-5382-6138}\inst{\ref{aff28}}
\and S.~Maurogordato\inst{\ref{aff98}}
\and E.~Medinaceli\orcid{0000-0002-4040-7783}\inst{\ref{aff29}}
\and S.~Mei\orcid{0000-0002-2849-559X}\inst{\ref{aff99},\ref{aff100}}
\and M.~Melchior\inst{\ref{aff95}}
\and M.~Meneghetti\orcid{0000-0003-1225-7084}\inst{\ref{aff29},\ref{aff33}}
\and E.~Merlin\orcid{0000-0001-6870-8900}\inst{\ref{aff57}}
\and G.~Meylan\inst{\ref{aff101}}
\and A.~Mora\orcid{0000-0002-1922-8529}\inst{\ref{aff102}}
\and M.~Moresco\orcid{0000-0002-7616-7136}\inst{\ref{aff97},\ref{aff29}}
\and L.~Moscardini\orcid{0000-0002-3473-6716}\inst{\ref{aff97},\ref{aff29},\ref{aff33}}
\and C.~Neissner\orcid{0000-0001-8524-4968}\inst{\ref{aff103},\ref{aff53}}
\and R.~C.~Nichol\orcid{0000-0003-0939-6518}\inst{\ref{aff25}}
\and C.~Padilla\orcid{0000-0001-7951-0166}\inst{\ref{aff103}}
\and S.~Paltani\orcid{0000-0002-8108-9179}\inst{\ref{aff71}}
\and F.~Pasian\orcid{0000-0002-4869-3227}\inst{\ref{aff18}}
\and K.~Pedersen\inst{\ref{aff104}}
\and W.~J.~Percival\orcid{0000-0002-0644-5727}\inst{\ref{aff105},\ref{aff106},\ref{aff107}}
\and V.~Pettorino\inst{\ref{aff16}}
\and S.~Pires\orcid{0000-0002-0249-2104}\inst{\ref{aff10}}
\and G.~Polenta\orcid{0000-0003-4067-9196}\inst{\ref{aff40}}
\and M.~Poncet\inst{\ref{aff11}}
\and L.~A.~Popa\inst{\ref{aff108}}
\and L.~Pozzetti\orcid{0000-0001-7085-0412}\inst{\ref{aff29}}
\and G.~D.~Racca\inst{\ref{aff16},\ref{aff12}}
\and F.~Raison\orcid{0000-0002-7819-6918}\inst{\ref{aff37}}
\and R.~Rebolo\orcid{0000-0003-3767-7085}\inst{\ref{aff61},\ref{aff109},\ref{aff110}}
\and A.~Renzi\orcid{0000-0001-9856-1970}\inst{\ref{aff111},\ref{aff73}}
\and J.~Rhodes\orcid{0000-0002-4485-8549}\inst{\ref{aff81}}
\and G.~Riccio\inst{\ref{aff44}}
\and E.~Romelli\orcid{0000-0003-3069-9222}\inst{\ref{aff18}}
\and M.~Roncarelli\orcid{0000-0001-9587-7822}\inst{\ref{aff29}}
\and E.~Rossetti\orcid{0000-0003-0238-4047}\inst{\ref{aff32}}
\and B.~Rusholme\orcid{0000-0001-7648-4142}\inst{\ref{aff27}}
\and R.~Saglia\orcid{0000-0003-0378-7032}\inst{\ref{aff38},\ref{aff37}}
\and Z.~Sakr\orcid{0000-0002-4823-3757}\inst{\ref{aff112},\ref{aff113},\ref{aff114}}
\and A.~G.~S\'anchez\orcid{0000-0003-1198-831X}\inst{\ref{aff37}}
\and D.~Sapone\orcid{0000-0001-7089-4503}\inst{\ref{aff115}}
\and B.~Sartoris\orcid{0000-0003-1337-5269}\inst{\ref{aff38},\ref{aff18}}
\and J.~A.~Schewtschenko\inst{\ref{aff62}}
\and P.~Schneider\orcid{0000-0001-8561-2679}\inst{\ref{aff5}}
\and T.~Schrabback\orcid{0000-0002-6987-7834}\inst{\ref{aff116}}
\and A.~Secroun\orcid{0000-0003-0505-3710}\inst{\ref{aff74}}
\and G.~Seidel\orcid{0000-0003-2907-353X}\inst{\ref{aff17}}
\and M.~Seiffert\orcid{0000-0002-7536-9393}\inst{\ref{aff81}}
\and S.~Serrano\orcid{0000-0002-0211-2861}\inst{\ref{aff56},\ref{aff117},\ref{aff55}}
\and P.~Simon\inst{\ref{aff5}}
\and C.~Sirignano\orcid{0000-0002-0995-7146}\inst{\ref{aff111},\ref{aff73}}
\and G.~Sirri\orcid{0000-0003-2626-2853}\inst{\ref{aff33}}
\and A.~Spurio~Mancini\orcid{0000-0001-5698-0990}\inst{\ref{aff118}}
\and L.~Stanco\orcid{0000-0002-9706-5104}\inst{\ref{aff73}}
\and J.~Steinwagner\orcid{0000-0001-7443-1047}\inst{\ref{aff37}}
\and P.~Tallada-Cresp\'{i}\orcid{0000-0002-1336-8328}\inst{\ref{aff52},\ref{aff53}}
\and D.~Tavagnacco\orcid{0000-0001-7475-9894}\inst{\ref{aff18}}
\and A.~N.~Taylor\inst{\ref{aff62}}
\and H.~I.~Teplitz\orcid{0000-0002-7064-5424}\inst{\ref{aff28}}
\and I.~Tereno\inst{\ref{aff69},\ref{aff119}}
\and N.~Tessore\orcid{0000-0002-9696-7931}\inst{\ref{aff86}}
\and S.~Toft\orcid{0000-0003-3631-7176}\inst{\ref{aff120},\ref{aff121}}
\and R.~Toledo-Moreo\orcid{0000-0002-2997-4859}\inst{\ref{aff122}}
\and F.~Torradeflot\orcid{0000-0003-1160-1517}\inst{\ref{aff53},\ref{aff52}}
\and I.~Tutusaus\orcid{0000-0002-3199-0399}\inst{\ref{aff113}}
\and E.~A.~Valentijn\inst{\ref{aff35}}
\and L.~Valenziano\orcid{0000-0002-1170-0104}\inst{\ref{aff29},\ref{aff75}}
\and J.~Valiviita\orcid{0000-0001-6225-3693}\inst{\ref{aff89},\ref{aff90}}
\and T.~Vassallo\orcid{0000-0001-6512-6358}\inst{\ref{aff38},\ref{aff18}}
\and G.~Verdoes~Kleijn\orcid{0000-0001-5803-2580}\inst{\ref{aff35}}
\and A.~Veropalumbo\orcid{0000-0003-2387-1194}\inst{\ref{aff26},\ref{aff42},\ref{aff41}}
\and Y.~Wang\orcid{0000-0002-4749-2984}\inst{\ref{aff28}}
\and J.~Weller\orcid{0000-0002-8282-2010}\inst{\ref{aff38},\ref{aff37}}
\and G.~Zamorani\orcid{0000-0002-2318-301X}\inst{\ref{aff29}}
\and F.~M.~Zerbi\inst{\ref{aff26}}
\and I.~A.~Zinchenko\orcid{0000-0002-2944-2449}\inst{\ref{aff38}}
\and E.~Zucca\orcid{0000-0002-5845-8132}\inst{\ref{aff29}}
\and V.~Allevato\orcid{0000-0001-7232-5152}\inst{\ref{aff44}}
\and M.~Ballardini\orcid{0000-0003-4481-3559}\inst{\ref{aff123},\ref{aff124},\ref{aff29}}
\and M.~Bolzonella\orcid{0000-0003-3278-4607}\inst{\ref{aff29}}
\and E.~Bozzo\orcid{0000-0002-8201-1525}\inst{\ref{aff71}}
\and C.~Burigana\orcid{0000-0002-3005-5796}\inst{\ref{aff125},\ref{aff75}}
\and R.~Cabanac\orcid{0000-0001-6679-2600}\inst{\ref{aff113}}
\and M.~Calabrese\orcid{0000-0002-2637-2422}\inst{\ref{aff126},\ref{aff51}}
\and A.~Cappi\inst{\ref{aff29},\ref{aff98}}
\and D.~Di~Ferdinando\inst{\ref{aff33}}
\and J.~A.~Escartin~Vigo\inst{\ref{aff37}}
\and G.~Fabbian\orcid{0000-0002-3255-4695}\inst{\ref{aff127}}
\and L.~Gabarra\orcid{0000-0002-8486-8856}\inst{\ref{aff15}}
\and M.~Huertas-Company\orcid{0000-0002-1416-8483}\inst{\ref{aff61},\ref{aff128},\ref{aff129},\ref{aff130}}
\and J.~Mart\'{i}n-Fleitas\orcid{0000-0002-8594-569X}\inst{\ref{aff102}}
\and S.~Matthew\orcid{0000-0001-8448-1697}\inst{\ref{aff62}}
\and N.~Mauri\orcid{0000-0001-8196-1548}\inst{\ref{aff60},\ref{aff33}}
\and R.~B.~Metcalf\orcid{0000-0003-3167-2574}\inst{\ref{aff97},\ref{aff29}}
\and A.~Pezzotta\orcid{0000-0003-0726-2268}\inst{\ref{aff131},\ref{aff37}}
\and M.~P\"ontinen\orcid{0000-0001-5442-2530}\inst{\ref{aff89}}
\and C.~Porciani\orcid{0000-0002-7797-2508}\inst{\ref{aff5}}
\and I.~Risso\orcid{0000-0003-2525-7761}\inst{\ref{aff132}}
\and V.~Scottez\inst{\ref{aff14},\ref{aff133}}
\and M.~Sereno\orcid{0000-0003-0302-0325}\inst{\ref{aff29},\ref{aff33}}
\and M.~Tenti\orcid{0000-0002-4254-5901}\inst{\ref{aff33}}
\and M.~Viel\orcid{0000-0002-2642-5707}\inst{\ref{aff19},\ref{aff18},\ref{aff31},\ref{aff30},\ref{aff134}}
\and M.~Wiesmann\orcid{0009-0000-8199-5860}\inst{\ref{aff80}}
\and Y.~Akrami\orcid{0000-0002-2407-7956}\inst{\ref{aff135},\ref{aff136}}
\and I.~T.~Andika\orcid{0000-0001-6102-9526}\inst{\ref{aff137},\ref{aff138}}
\and S.~Anselmi\orcid{0000-0002-3579-9583}\inst{\ref{aff73},\ref{aff111},\ref{aff139}}
\and M.~Archidiacono\orcid{0000-0003-4952-9012}\inst{\ref{aff78},\ref{aff79}}
\and F.~Atrio-Barandela\orcid{0000-0002-2130-2513}\inst{\ref{aff140}}
\and C.~Benoist\inst{\ref{aff98}}
\and P.~Bergamini\orcid{0000-0003-1383-9414}\inst{\ref{aff78},\ref{aff29}}
\and D.~Bertacca\orcid{0000-0002-2490-7139}\inst{\ref{aff111},\ref{aff34},\ref{aff73}}
\and M.~Bethermin\orcid{0000-0002-3915-2015}\inst{\ref{aff22}}
\and L.~Bisigello\orcid{0000-0003-0492-4924}\inst{\ref{aff34}}
\and A.~Blanchard\orcid{0000-0001-8555-9003}\inst{\ref{aff113}}
\and L.~Blot\orcid{0000-0002-9622-7167}\inst{\ref{aff141},\ref{aff139}}
\and S.~Borgani\orcid{0000-0001-6151-6439}\inst{\ref{aff142},\ref{aff19},\ref{aff18},\ref{aff30},\ref{aff134}}
\and A.~S.~Borlaff\orcid{0000-0003-3249-4431}\inst{\ref{aff21},\ref{aff143}}
\and M.~L.~Brown\orcid{0000-0002-0370-8077}\inst{\ref{aff63}}
\and S.~Bruton\orcid{0000-0002-6503-5218}\inst{\ref{aff144}}
\and A.~Calabro\orcid{0000-0003-2536-1614}\inst{\ref{aff57}}
\and B.~Camacho~Quevedo\orcid{0000-0002-8789-4232}\inst{\ref{aff56},\ref{aff55}}
\and F.~Caro\inst{\ref{aff57}}
\and C.~S.~Carvalho\inst{\ref{aff119}}
\and T.~Castro\orcid{0000-0002-6292-3228}\inst{\ref{aff18},\ref{aff30},\ref{aff19},\ref{aff134}}
\and F.~Cogato\orcid{0000-0003-4632-6113}\inst{\ref{aff97},\ref{aff29}}
\and S.~Conseil\orcid{0000-0002-3657-4191}\inst{\ref{aff65}}
\and A.~R.~Cooray\orcid{0000-0002-3892-0190}\inst{\ref{aff145}}
\and O.~Cucciati\orcid{0000-0002-9336-7551}\inst{\ref{aff29}}
\and S.~Davini\orcid{0000-0003-3269-1718}\inst{\ref{aff42}}
\and F.~De~Paolis\orcid{0000-0001-6460-7563}\inst{\ref{aff146},\ref{aff147},\ref{aff148}}
\and G.~Desprez\orcid{0000-0001-8325-1742}\inst{\ref{aff35}}
\and A.~D\'iaz-S\'anchez\orcid{0000-0003-0748-4768}\inst{\ref{aff149}}
\and J.~J.~Diaz\inst{\ref{aff128}}
\and S.~Di~Domizio\orcid{0000-0003-2863-5895}\inst{\ref{aff41},\ref{aff42}}
\and J.~M.~Diego\orcid{0000-0001-9065-3926}\inst{\ref{aff150}}
\and P.-A.~Duc\orcid{0000-0003-3343-6284}\inst{\ref{aff22}}
\and A.~Enia\orcid{0000-0002-0200-2857}\inst{\ref{aff32},\ref{aff29}}
\and Y.~Fang\inst{\ref{aff38}}
\and A.~M.~N.~Ferguson\inst{\ref{aff62}}
\and A.~G.~Ferrari\orcid{0009-0005-5266-4110}\inst{\ref{aff33}}
\and A.~Finoguenov\orcid{0000-0002-4606-5403}\inst{\ref{aff89}}
\and A.~Fontana\orcid{0000-0003-3820-2823}\inst{\ref{aff57}}
\and A.~Franco\orcid{0000-0002-4761-366X}\inst{\ref{aff147},\ref{aff146},\ref{aff148}}
\and K.~Ganga\orcid{0000-0001-8159-8208}\inst{\ref{aff99}}
\and J.~Garc\'ia-Bellido\orcid{0000-0002-9370-8360}\inst{\ref{aff135}}
\and T.~Gasparetto\orcid{0000-0002-7913-4866}\inst{\ref{aff18}}
\and V.~Gautard\inst{\ref{aff151}}
\and E.~Gaztanaga\orcid{0000-0001-9632-0815}\inst{\ref{aff55},\ref{aff56},\ref{aff152}}
\and F.~Giacomini\orcid{0000-0002-3129-2814}\inst{\ref{aff33}}
\and F.~Gianotti\orcid{0000-0003-4666-119X}\inst{\ref{aff29}}
\and A.~H.~Gonzalez\orcid{0000-0002-0933-8601}\inst{\ref{aff153}}
\and G.~Gozaliasl\orcid{0000-0002-0236-919X}\inst{\ref{aff154},\ref{aff89}}
\and A.~Gregorio\orcid{0000-0003-4028-8785}\inst{\ref{aff142},\ref{aff18},\ref{aff30}}
\and M.~Guidi\orcid{0000-0001-9408-1101}\inst{\ref{aff32},\ref{aff29}}
\and C.~M.~Gutierrez\orcid{0000-0001-7854-783X}\inst{\ref{aff155}}
\and A.~Hall\orcid{0000-0002-3139-8651}\inst{\ref{aff62}}
\and W.~G.~Hartley\inst{\ref{aff71}}
\and C.~Hern\'andez-Monteagudo\orcid{0000-0001-5471-9166}\inst{\ref{aff110},\ref{aff61}}
\and H.~Hildebrandt\orcid{0000-0002-9814-3338}\inst{\ref{aff156}}
\and J.~Hjorth\orcid{0000-0002-4571-2306}\inst{\ref{aff104}}
\and J.~J.~E.~Kajava\orcid{0000-0002-3010-8333}\inst{\ref{aff157},\ref{aff158}}
\and Y.~Kang\orcid{0009-0000-8588-7250}\inst{\ref{aff71}}
\and V.~Kansal\orcid{0000-0002-4008-6078}\inst{\ref{aff159},\ref{aff160}}
\and D.~Karagiannis\orcid{0000-0002-4927-0816}\inst{\ref{aff123},\ref{aff161}}
\and K.~Kiiveri\inst{\ref{aff87}}
\and C.~C.~Kirkpatrick\inst{\ref{aff87}}
\and S.~Kruk\orcid{0000-0001-8010-8879}\inst{\ref{aff24}}
\and J.~Le~Graet\orcid{0000-0001-6523-7971}\inst{\ref{aff74}}
\and L.~Legrand\orcid{0000-0003-0610-5252}\inst{\ref{aff162},\ref{aff163}}
\and M.~Lembo\orcid{0000-0002-5271-5070}\inst{\ref{aff123},\ref{aff124}}
\and F.~Lepori\orcid{0009-0000-5061-7138}\inst{\ref{aff164}}
\and G.~F.~Lesci\orcid{0000-0002-4607-2830}\inst{\ref{aff97},\ref{aff29}}
\and J.~Lesgourgues\orcid{0000-0001-7627-353X}\inst{\ref{aff54}}
\and L.~Leuzzi\orcid{0009-0006-4479-7017}\inst{\ref{aff97},\ref{aff29}}
\and T.~I.~Liaudat\orcid{0000-0002-9104-314X}\inst{\ref{aff165}}
\and S.~J.~Liu\orcid{0000-0001-7680-2139}\inst{\ref{aff72}}
\and A.~Loureiro\orcid{0000-0002-4371-0876}\inst{\ref{aff166},\ref{aff167}}
\and J.~Macias-Perez\orcid{0000-0002-5385-2763}\inst{\ref{aff168}}
\and G.~Maggio\orcid{0000-0003-4020-4836}\inst{\ref{aff18}}
\and M.~Magliocchetti\orcid{0000-0001-9158-4838}\inst{\ref{aff72}}
\and F.~Mannucci\orcid{0000-0002-4803-2381}\inst{\ref{aff169}}
\and R.~Maoli\orcid{0000-0002-6065-3025}\inst{\ref{aff170},\ref{aff57}}
\and C.~J.~A.~P.~Martins\orcid{0000-0002-4886-9261}\inst{\ref{aff171},\ref{aff45}}
\and L.~Maurin\orcid{0000-0002-8406-0857}\inst{\ref{aff23}}
\and C.~J.~R.~McPartland\orcid{0000-0003-0639-025X}\inst{\ref{aff84},\ref{aff121}}
\and M.~Miluzio\inst{\ref{aff24},\ref{aff172}}
\and P.~Monaco\orcid{0000-0003-2083-7564}\inst{\ref{aff142},\ref{aff18},\ref{aff30},\ref{aff19}}
\and A.~Montoro\orcid{0000-0003-4730-8590}\inst{\ref{aff55},\ref{aff56}}
\and C.~Moretti\orcid{0000-0003-3314-8936}\inst{\ref{aff31},\ref{aff134},\ref{aff18},\ref{aff19},\ref{aff30}}
\and G.~Morgante\inst{\ref{aff29}}
\and C.~Murray\inst{\ref{aff99}}
\and S.~Nadathur\orcid{0000-0001-9070-3102}\inst{\ref{aff152}}
\and K.~Naidoo\orcid{0000-0002-9182-1802}\inst{\ref{aff152}}
\and A.~Navarro-Alsina\orcid{0000-0002-3173-2592}\inst{\ref{aff5}}
\and S.~Nesseris\orcid{0000-0002-0567-0324}\inst{\ref{aff135}}
\and F.~Passalacqua\orcid{0000-0002-8606-4093}\inst{\ref{aff111},\ref{aff73}}
\and L.~Patrizii\inst{\ref{aff33}}
\and A.~Pisani\orcid{0000-0002-6146-4437}\inst{\ref{aff74},\ref{aff173}}
\and D.~Potter\orcid{0000-0002-0757-5195}\inst{\ref{aff164}}
\and S.~Quai\orcid{0000-0002-0449-8163}\inst{\ref{aff97},\ref{aff29}}
\and M.~Radovich\orcid{0000-0002-3585-866X}\inst{\ref{aff34}}
\and P.~Reimberg\orcid{0000-0003-3410-0280}\inst{\ref{aff14}}
\and P.-F.~Rocci\inst{\ref{aff23}}
\and G.~Rodighiero\orcid{0000-0002-9415-2296}\inst{\ref{aff111},\ref{aff34}}
\and R.~P.~Rollins\orcid{0000-0003-1291-1023}\inst{\ref{aff62}}
\and S.~Sacquegna\orcid{0000-0002-8433-6630}\inst{\ref{aff146},\ref{aff147},\ref{aff148}}
\and M.~Sahl\'en\orcid{0000-0003-0973-4804}\inst{\ref{aff174}}
\and D.~B.~Sanders\orcid{0000-0002-1233-9998}\inst{\ref{aff59}}
\and E.~Sarpa\orcid{0000-0002-1256-655X}\inst{\ref{aff31},\ref{aff134},\ref{aff30}}
\and C.~Scarlata\orcid{0000-0002-9136-8876}\inst{\ref{aff175}}
\and A.~Schneider\orcid{0000-0001-7055-8104}\inst{\ref{aff164}}
\and M.~Schultheis\inst{\ref{aff98}}
\and D.~Sciotti\orcid{0009-0008-4519-2620}\inst{\ref{aff57},\ref{aff96}}
\and E.~Sellentin\inst{\ref{aff176},\ref{aff12}}
\and F.~Shankar\orcid{0000-0001-8973-5051}\inst{\ref{aff177}}
\and A.~Silvestri\orcid{0000-0001-6904-5061}\inst{\ref{aff50}}
\and K.~Tanidis\orcid{0000-0001-9843-5130}\inst{\ref{aff15}}
\and C.~Tao\orcid{0000-0001-7961-8177}\inst{\ref{aff74}}
\and G.~Testera\inst{\ref{aff42}}
\and M.~Tewes\orcid{0000-0002-1155-8689}\inst{\ref{aff5}}
\and R.~Teyssier\orcid{0000-0001-7689-0933}\inst{\ref{aff173}}
\and S.~Tosi\orcid{0000-0002-7275-9193}\inst{\ref{aff41},\ref{aff132}}
\and A.~Troja\orcid{0000-0003-0239-4595}\inst{\ref{aff111},\ref{aff73}}
\and M.~Tucci\inst{\ref{aff71}}
\and C.~Valieri\inst{\ref{aff33}}
\and A.~Venhola\orcid{0000-0001-6071-4564}\inst{\ref{aff178}}
\and D.~Vergani\orcid{0000-0003-0898-2216}\inst{\ref{aff29}}
\and G.~Verza\orcid{0000-0002-1886-8348}\inst{\ref{aff179}}
\and P.~Vielzeuf\orcid{0000-0003-2035-9339}\inst{\ref{aff74}}
\and J.~R.~Weaver\orcid{0000-0003-1614-196X}\inst{\ref{aff180}}
\and D.~Scott\orcid{0000-0002-6878-9840}\inst{\ref{aff181}}}
										   
\institute{Institut d'Astrophysique de Paris, UMR 7095, CNRS, and Sorbonne Universit\'e, 98 bis boulevard Arago, 75014 Paris, France\label{aff1}
\and
Mullard Space Science Laboratory, University College London, Holmbury St Mary, Dorking, Surrey RH5 6NT, UK\label{aff2}
\and
Department of Physics, Centre for Extragalactic Astronomy, Durham University, South Road, Durham, DH1 3LE, UK\label{aff3}
\and
Department of Physics, Institute for Computational Cosmology, Durham University, South Road, Durham, DH1 3LE, UK\label{aff4}
\and
Universit\"at Bonn, Argelander-Institut f\"ur Astronomie, Auf dem H\"ugel 71, 53121 Bonn, Germany\label{aff5}
\and
School of Mathematics, Statistics and Physics, Newcastle University, Herschel Building, Newcastle-upon-Tyne, NE1 7RU, UK\label{aff6}
\and
Centre for Electronic Imaging, Open University, Walton Hall, Milton Keynes, MK7~6AA, UK\label{aff7}
\and
Institute of Astronomy, University of Cambridge, Madingley Road, Cambridge CB3 0HA, UK\label{aff8}
\and
Centro de Estudios de F\'isica del Cosmos de Arag\'on (CEFCA), Plaza San Juan, 1, planta 2, 44001, Teruel, Spain\label{aff9}
\and
Universit\'e Paris-Saclay, Universit\'e Paris Cit\'e, CEA, CNRS, AIM, 91191, Gif-sur-Yvette, France\label{aff10}
\and
Centre National d'Etudes Spatiales -- Centre spatial de Toulouse, 18 avenue Edouard Belin, 31401 Toulouse Cedex 9, France\label{aff11}
\and
Leiden Observatory, Leiden University, Einsteinweg 55, 2333 CC Leiden, The Netherlands\label{aff12}
\and
Ernst-Reuter-Str. 4e, 31224 Peine, Germany\label{aff13}
\and
Institut d'Astrophysique de Paris, 98bis Boulevard Arago, 75014, Paris, France\label{aff14}
\and
Department of Physics, Oxford University, Keble Road, Oxford OX1 3RH, UK\label{aff15}
\and
European Space Agency/ESTEC, Keplerlaan 1, 2201 AZ Noordwijk, The Netherlands\label{aff16}
\and
Max-Planck-Institut f\"ur Astronomie, K\"onigstuhl 17, 69117 Heidelberg, Germany\label{aff17}
\and
INAF-Osservatorio Astronomico di Trieste, Via G. B. Tiepolo 11, 34143 Trieste, Italy\label{aff18}
\and
IFPU, Institute for Fundamental Physics of the Universe, via Beirut 2, 34151 Trieste, Italy\label{aff19}
\and
Cavendish Laboratory, University of Cambridge, JJ Thomson Avenue, Cambridge, CB3 0HE, UK\label{aff20}
\and
NASA Ames Research Center, Moffett Field, CA 94035, USA\label{aff21}
\and
Universit\'e de Strasbourg, CNRS, Observatoire astronomique de Strasbourg, UMR 7550, 67000 Strasbourg, France\label{aff22}
\and
Universit\'e Paris-Saclay, CNRS, Institut d'astrophysique spatiale, 91405, Orsay, France\label{aff23}
\and
ESAC/ESA, Camino Bajo del Castillo, s/n., Urb. Villafranca del Castillo, 28692 Villanueva de la Ca\~nada, Madrid, Spain\label{aff24}
\and
School of Mathematics and Physics, University of Surrey, Guildford, Surrey, GU2 7XH, UK\label{aff25}
\and
INAF-Osservatorio Astronomico di Brera, Via Brera 28, 20122 Milano, Italy\label{aff26}
\and
Caltech/IPAC, 1200 E. California Blvd., Pasadena, CA 91125, USA\label{aff27}
\and
Infrared Processing and Analysis Center, California Institute of Technology, Pasadena, CA 91125, USA\label{aff28}
\and
INAF-Osservatorio di Astrofisica e Scienza dello Spazio di Bologna, Via Piero Gobetti 93/3, 40129 Bologna, Italy\label{aff29}
\and
INFN, Sezione di Trieste, Via Valerio 2, 34127 Trieste TS, Italy\label{aff30}
\and
SISSA, International School for Advanced Studies, Via Bonomea 265, 34136 Trieste TS, Italy\label{aff31}
\and
Dipartimento di Fisica e Astronomia, Universit\`a di Bologna, Via Gobetti 93/2, 40129 Bologna, Italy\label{aff32}
\and
INFN-Sezione di Bologna, Viale Berti Pichat 6/2, 40127 Bologna, Italy\label{aff33}
\and
INAF-Osservatorio Astronomico di Padova, Via dell'Osservatorio 5, 35122 Padova, Italy\label{aff34}
\and
Kapteyn Astronomical Institute, University of Groningen, PO Box 800, 9700 AV Groningen, The Netherlands\label{aff35}
\and
ATG Europe BV, Huygensstraat 34, 2201 DK Noordwijk, The Netherlands\label{aff36}
\and
Max Planck Institute for Extraterrestrial Physics, Giessenbachstr. 1, 85748 Garching, Germany\label{aff37}
\and
Universit\"ats-Sternwarte M\"unchen, Fakult\"at f\"ur Physik, Ludwig-Maximilians-Universit\"at M\"unchen, Scheinerstrasse 1, 81679 M\"unchen, Germany\label{aff38}
\and
Institut de Physique Th\'eorique, CEA, CNRS, Universit\'e Paris-Saclay 91191 Gif-sur-Yvette Cedex, France\label{aff39}
\and
Space Science Data Center, Italian Space Agency, via del Politecnico snc, 00133 Roma, Italy\label{aff40}
\and
Dipartimento di Fisica, Universit\`a di Genova, Via Dodecaneso 33, 16146, Genova, Italy\label{aff41}
\and
INFN-Sezione di Genova, Via Dodecaneso 33, 16146, Genova, Italy\label{aff42}
\and
Department of Physics "E. Pancini", University Federico II, Via Cinthia 6, 80126, Napoli, Italy\label{aff43}
\and
INAF-Osservatorio Astronomico di Capodimonte, Via Moiariello 16, 80131 Napoli, Italy\label{aff44}
\and
Instituto de Astrof\'isica e Ci\^encias do Espa\c{c}o, Universidade do Porto, CAUP, Rua das Estrelas, PT4150-762 Porto, Portugal\label{aff45}
\and
Faculdade de Ci\^encias da Universidade do Porto, Rua do Campo de Alegre, 4150-007 Porto, Portugal\label{aff46}
\and
Dipartimento di Fisica, Universit\`a degli Studi di Torino, Via P. Giuria 1, 10125 Torino, Italy\label{aff47}
\and
INFN-Sezione di Torino, Via P. Giuria 1, 10125 Torino, Italy\label{aff48}
\and
INAF-Osservatorio Astrofisico di Torino, Via Osservatorio 20, 10025 Pino Torinese (TO), Italy\label{aff49}
\and
Institute Lorentz, Leiden University, Niels Bohrweg 2, 2333 CA Leiden, The Netherlands\label{aff50}
\and
INAF-IASF Milano, Via Alfonso Corti 12, 20133 Milano, Italy\label{aff51}
\and
Centro de Investigaciones Energ\'eticas, Medioambientales y Tecnol\'ogicas (CIEMAT), Avenida Complutense 40, 28040 Madrid, Spain\label{aff52}
\and
Port d'Informaci\'{o} Cient\'{i}fica, Campus UAB, C. Albareda s/n, 08193 Bellaterra (Barcelona), Spain\label{aff53}
\and
Institute for Theoretical Particle Physics and Cosmology (TTK), RWTH Aachen University, 52056 Aachen, Germany\label{aff54}
\and
Institute of Space Sciences (ICE, CSIC), Campus UAB, Carrer de Can Magrans, s/n, 08193 Barcelona, Spain\label{aff55}
\and
Institut d'Estudis Espacials de Catalunya (IEEC),  Edifici RDIT, Campus UPC, 08860 Castelldefels, Barcelona, Spain\label{aff56}
\and
INAF-Osservatorio Astronomico di Roma, Via Frascati 33, 00078 Monteporzio Catone, Italy\label{aff57}
\and
INFN section of Naples, Via Cinthia 6, 80126, Napoli, Italy\label{aff58}
\and
Institute for Astronomy, University of Hawaii, 2680 Woodlawn Drive, Honolulu, HI 96822, USA\label{aff59}
\and
Dipartimento di Fisica e Astronomia "Augusto Righi" - Alma Mater Studiorum Universit\`a di Bologna, Viale Berti Pichat 6/2, 40127 Bologna, Italy\label{aff60}
\and
Instituto de Astrof\'{\i}sica de Canarias, V\'{\i}a L\'actea, 38205 La Laguna, Tenerife, Spain\label{aff61}
\and
Institute for Astronomy, University of Edinburgh, Royal Observatory, Blackford Hill, Edinburgh EH9 3HJ, UK\label{aff62}
\and
Jodrell Bank Centre for Astrophysics, Department of Physics and Astronomy, University of Manchester, Oxford Road, Manchester M13 9PL, UK\label{aff63}
\and
European Space Agency/ESRIN, Largo Galileo Galilei 1, 00044 Frascati, Roma, Italy\label{aff64}
\and
Universit\'e Claude Bernard Lyon 1, CNRS/IN2P3, IP2I Lyon, UMR 5822, Villeurbanne, F-69100, France\label{aff65}
\and
Institut de Ci\`{e}ncies del Cosmos (ICCUB), Universitat de Barcelona (IEEC-UB), Mart\'{i} i Franqu\`{e}s 1, 08028 Barcelona, Spain\label{aff66}
\and
Instituci\'o Catalana de Recerca i Estudis Avan\c{c}ats (ICREA), Passeig de Llu\'{\i}s Companys 23, 08010 Barcelona, Spain\label{aff67}
\and
UCB Lyon 1, CNRS/IN2P3, IUF, IP2I Lyon, 4 rue Enrico Fermi, 69622 Villeurbanne, France\label{aff68}
\and
Departamento de F\'isica, Faculdade de Ci\^encias, Universidade de Lisboa, Edif\'icio C8, Campo Grande, PT1749-016 Lisboa, Portugal\label{aff69}
\and
Instituto de Astrof\'isica e Ci\^encias do Espa\c{c}o, Faculdade de Ci\^encias, Universidade de Lisboa, Campo Grande, 1749-016 Lisboa, Portugal\label{aff70}
\and
Department of Astronomy, University of Geneva, ch. d'Ecogia 16, 1290 Versoix, Switzerland\label{aff71}
\and
INAF-Istituto di Astrofisica e Planetologia Spaziali, via del Fosso del Cavaliere, 100, 00100 Roma, Italy\label{aff72}
\and
INFN-Padova, Via Marzolo 8, 35131 Padova, Italy\label{aff73}
\and
Aix-Marseille Universit\'e, CNRS/IN2P3, CPPM, Marseille, France\label{aff74}
\and
INFN-Bologna, Via Irnerio 46, 40126 Bologna, Italy\label{aff75}
\and
School of Physics, HH Wills Physics Laboratory, University of Bristol, Tyndall Avenue, Bristol, BS8 1TL, UK\label{aff76}
\and
FRACTAL S.L.N.E., calle Tulip\'an 2, Portal 13 1A, 28231, Las Rozas de Madrid, Spain\label{aff77}
\and
Dipartimento di Fisica "Aldo Pontremoli", Universit\`a degli Studi di Milano, Via Celoria 16, 20133 Milano, Italy\label{aff78}
\and
INFN-Sezione di Milano, Via Celoria 16, 20133 Milano, Italy\label{aff79}
\and
Institute of Theoretical Astrophysics, University of Oslo, P.O. Box 1029 Blindern, 0315 Oslo, Norway\label{aff80}
\and
Jet Propulsion Laboratory, California Institute of Technology, 4800 Oak Grove Drive, Pasadena, CA, 91109, USA\label{aff81}
\and
Felix Hormuth Engineering, Goethestr. 17, 69181 Leimen, Germany\label{aff82}
\and
Technical University of Denmark, Elektrovej 327, 2800 Kgs. Lyngby, Denmark\label{aff83}
\and
Cosmic Dawn Center (DAWN), Denmark\label{aff84}
\and
NASA Goddard Space Flight Center, Greenbelt, MD 20771, USA\label{aff85}
\and
Department of Physics and Astronomy, University College London, Gower Street, London WC1E 6BT, UK\label{aff86}
\and
Department of Physics and Helsinki Institute of Physics, Gustaf H\"allstr\"omin katu 2, 00014 University of Helsinki, Finland\label{aff87}
\and
Universit\'e de Gen\`eve, D\'epartement de Physique Th\'eorique and Centre for Astroparticle Physics, 24 quai Ernest-Ansermet, CH-1211 Gen\`eve 4, Switzerland\label{aff88}
\and
Department of Physics, P.O. Box 64, 00014 University of Helsinki, Finland\label{aff89}
\and
Helsinki Institute of Physics, Gustaf H{\"a}llstr{\"o}min katu 2, University of Helsinki, Helsinki, Finland\label{aff90}
\and
Centre de Calcul de l'IN2P3/CNRS, 21 avenue Pierre de Coubertin 69627 Villeurbanne Cedex, France\label{aff91}
\and
Laboratoire d'etude de l'Univers et des phenomenes eXtremes, Observatoire de Paris, Universit\'e PSL, Sorbonne Universit\'e, CNRS, 92190 Meudon, France\label{aff92}
\and
Aix-Marseille Universit\'e, CNRS, CNES, LAM, Marseille, France\label{aff93}
\and
SKA Observatory, Jodrell Bank, Lower Withington, Macclesfield, Cheshire SK11 9FT, UK\label{aff94}
\and
University of Applied Sciences and Arts of Northwestern Switzerland, School of Engineering, 5210 Windisch, Switzerland\label{aff95}
\and
INFN-Sezione di Roma, Piazzale Aldo Moro, 2 - c/o Dipartimento di Fisica, Edificio G. Marconi, 00185 Roma, Italy\label{aff96}
\and
Dipartimento di Fisica e Astronomia "Augusto Righi" - Alma Mater Studiorum Universit\`a di Bologna, via Piero Gobetti 93/2, 40129 Bologna, Italy\label{aff97}
\and
Universit\'e C\^{o}te d'Azur, Observatoire de la C\^{o}te d'Azur, CNRS, Laboratoire Lagrange, Bd de l'Observatoire, CS 34229, 06304 Nice cedex 4, France\label{aff98}
\and
Universit\'e Paris Cit\'e, CNRS, Astroparticule et Cosmologie, 75013 Paris, France\label{aff99}
\and
CNRS-UCB International Research Laboratory, Centre Pierre Binetruy, IRL2007, CPB-IN2P3, Berkeley, USA\label{aff100}
\and
Institute of Physics, Laboratory of Astrophysics, Ecole Polytechnique F\'ed\'erale de Lausanne (EPFL), Observatoire de Sauverny, 1290 Versoix, Switzerland\label{aff101}
\and
Aurora Technology for European Space Agency (ESA), Camino bajo del Castillo, s/n, Urbanizacion Villafranca del Castillo, Villanueva de la Ca\~nada, 28692 Madrid, Spain\label{aff102}
\and
Institut de F\'{i}sica d'Altes Energies (IFAE), The Barcelona Institute of Science and Technology, Campus UAB, 08193 Bellaterra (Barcelona), Spain\label{aff103}
\and
DARK, Niels Bohr Institute, University of Copenhagen, Jagtvej 155, 2200 Copenhagen, Denmark\label{aff104}
\and
Waterloo Centre for Astrophysics, University of Waterloo, Waterloo, Ontario N2L 3G1, Canada\label{aff105}
\and
Department of Physics and Astronomy, University of Waterloo, Waterloo, Ontario N2L 3G1, Canada\label{aff106}
\and
Perimeter Institute for Theoretical Physics, Waterloo, Ontario N2L 2Y5, Canada\label{aff107}
\and
Institute of Space Science, Str. Atomistilor, nr. 409 M\u{a}gurele, Ilfov, 077125, Romania\label{aff108}
\and
Consejo Superior de Investigaciones Cientificas, Calle Serrano 117, 28006 Madrid, Spain\label{aff109}
\and
Universidad de La Laguna, Departamento de Astrof\'{\i}sica, 38206 La Laguna, Tenerife, Spain\label{aff110}
\and
Dipartimento di Fisica e Astronomia "G. Galilei", Universit\`a di Padova, Via Marzolo 8, 35131 Padova, Italy\label{aff111}
\and
Institut f\"ur Theoretische Physik, University of Heidelberg, Philosophenweg 16, 69120 Heidelberg, Germany\label{aff112}
\and
Institut de Recherche en Astrophysique et Plan\'etologie (IRAP), Universit\'e de Toulouse, CNRS, UPS, CNES, 14 Av. Edouard Belin, 31400 Toulouse, France\label{aff113}
\and
Universit\'e St Joseph; Faculty of Sciences, Beirut, Lebanon\label{aff114}
\and
Departamento de F\'isica, FCFM, Universidad de Chile, Blanco Encalada 2008, Santiago, Chile\label{aff115}
\and
Universit\"at Innsbruck, Institut f\"ur Astro- und Teilchenphysik, Technikerstr. 25/8, 6020 Innsbruck, Austria\label{aff116}
\and
Satlantis, University Science Park, Sede Bld 48940, Leioa-Bilbao, Spain\label{aff117}
\and
Department of Physics, Royal Holloway, University of London, TW20 0EX, UK\label{aff118}
\and
Instituto de Astrof\'isica e Ci\^encias do Espa\c{c}o, Faculdade de Ci\^encias, Universidade de Lisboa, Tapada da Ajuda, 1349-018 Lisboa, Portugal\label{aff119}
\and
Cosmic Dawn Center (DAWN)\label{aff120}
\and
Niels Bohr Institute, University of Copenhagen, Jagtvej 128, 2200 Copenhagen, Denmark\label{aff121}
\and
Universidad Polit\'ecnica de Cartagena, Departamento de Electr\'onica y Tecnolog\'ia de Computadoras,  Plaza del Hospital 1, 30202 Cartagena, Spain\label{aff122}
\and
Dipartimento di Fisica e Scienze della Terra, Universit\`a degli Studi di Ferrara, Via Giuseppe Saragat 1, 44122 Ferrara, Italy\label{aff123}
\and
Istituto Nazionale di Fisica Nucleare, Sezione di Ferrara, Via Giuseppe Saragat 1, 44122 Ferrara, Italy\label{aff124}
\and
INAF, Istituto di Radioastronomia, Via Piero Gobetti 101, 40129 Bologna, Italy\label{aff125}
\and
Astronomical Observatory of the Autonomous Region of the Aosta Valley (OAVdA), Loc. Lignan 39, I-11020, Nus (Aosta Valley), Italy\label{aff126}
\and
School of Physics and Astronomy, Cardiff University, The Parade, Cardiff, CF24 3AA, UK\label{aff127}
\and
Instituto de Astrof\'isica de Canarias (IAC); Departamento de Astrof\'isica, Universidad de La Laguna (ULL), 38200, La Laguna, Tenerife, Spain\label{aff128}
\and
Universit\'e PSL, Observatoire de Paris, Sorbonne Universit\'e, CNRS, LERMA, 75014, Paris, France\label{aff129}
\and
Universit\'e Paris-Cit\'e, 5 Rue Thomas Mann, 75013, Paris, France\label{aff130}
\and
INAF - Osservatorio Astronomico di Brera, via Emilio Bianchi 46, 23807 Merate, Italy\label{aff131}
\and
INAF-Osservatorio Astronomico di Brera, Via Brera 28, 20122 Milano, Italy, and INFN-Sezione di Genova, Via Dodecaneso 33, 16146, Genova, Italy\label{aff132}
\and
ICL, Junia, Universit\'e Catholique de Lille, LITL, 59000 Lille, France\label{aff133}
\and
ICSC - Centro Nazionale di Ricerca in High Performance Computing, Big Data e Quantum Computing, Via Magnanelli 2, Bologna, Italy\label{aff134}
\and
Instituto de F\'isica Te\'orica UAM-CSIC, Campus de Cantoblanco, 28049 Madrid, Spain\label{aff135}
\and
CERCA/ISO, Department of Physics, Case Western Reserve University, 10900 Euclid Avenue, Cleveland, OH 44106, USA\label{aff136}
\and
Technical University of Munich, TUM School of Natural Sciences, Physics Department, James-Franck-Str.~1, 85748 Garching, Germany\label{aff137}
\and
Max-Planck-Institut f\"ur Astrophysik, Karl-Schwarzschild-Str.~1, 85748 Garching, Germany\label{aff138}
\and
Laboratoire Univers et Th\'eorie, Observatoire de Paris, Universit\'e PSL, Universit\'e Paris Cit\'e, CNRS, 92190 Meudon, France\label{aff139}
\and
Departamento de F{\'\i}sica Fundamental. Universidad de Salamanca. Plaza de la Merced s/n. 37008 Salamanca, Spain\label{aff140}
\and
Center for Data-Driven Discovery, Kavli IPMU (WPI), UTIAS, The University of Tokyo, Kashiwa, Chiba 277-8583, Japan\label{aff141}
\and
Dipartimento di Fisica - Sezione di Astronomia, Universit\`a di Trieste, Via Tiepolo 11, 34131 Trieste, Italy\label{aff142}
\and
Bay Area Environmental Research Institute, Moffett Field, California 94035, USA\label{aff143}
\and
California Institute of Technology, 1200 E California Blvd, Pasadena, CA 91125, USA\label{aff144}
\and
Department of Physics \& Astronomy, University of California Irvine, Irvine CA 92697, USA\label{aff145}
\and
Department of Mathematics and Physics E. De Giorgi, University of Salento, Via per Arnesano, CP-I93, 73100, Lecce, Italy\label{aff146}
\and
INFN, Sezione di Lecce, Via per Arnesano, CP-193, 73100, Lecce, Italy\label{aff147}
\and
INAF-Sezione di Lecce, c/o Dipartimento Matematica e Fisica, Via per Arnesano, 73100, Lecce, Italy\label{aff148}
\and
Departamento F\'isica Aplicada, Universidad Polit\'ecnica de Cartagena, Campus Muralla del Mar, 30202 Cartagena, Murcia, Spain\label{aff149}
\and
Instituto de F\'isica de Cantabria, Edificio Juan Jord\'a, Avenida de los Castros, 39005 Santander, Spain\label{aff150}
\and
CEA Saclay, DFR/IRFU, Service d'Astrophysique, Bat. 709, 91191 Gif-sur-Yvette, France\label{aff151}
\and
Institute of Cosmology and Gravitation, University of Portsmouth, Portsmouth PO1 3FX, UK\label{aff152}
\and
Department of Astronomy, University of Florida, Bryant Space Science Center, Gainesville, FL 32611, USA\label{aff153}
\and
Department of Computer Science, Aalto University, PO Box 15400, Espoo, FI-00 076, Finland\label{aff154}
\and
Instituto de Astrof\'\i sica de Canarias, c/ Via Lactea s/n, La Laguna 38200, Spain. Departamento de Astrof\'\i sica de la Universidad de La Laguna, Avda. Francisco Sanchez, La Laguna, 38200, Spain\label{aff155}
\and
Ruhr University Bochum, Faculty of Physics and Astronomy, Astronomical Institute (AIRUB), German Centre for Cosmological Lensing (GCCL), 44780 Bochum, Germany\label{aff156}
\and
Department of Physics and Astronomy, Vesilinnantie 5, 20014 University of Turku, Finland\label{aff157}
\and
Serco for European Space Agency (ESA), Camino bajo del Castillo, s/n, Urbanizacion Villafranca del Castillo, Villanueva de la Ca\~nada, 28692 Madrid, Spain\label{aff158}
\and
ARC Centre of Excellence for Dark Matter Particle Physics, Melbourne, Australia\label{aff159}
\and
Centre for Astrophysics \& Supercomputing, Swinburne University of Technology,  Hawthorn, Victoria 3122, Australia\label{aff160}
\and
Department of Physics and Astronomy, University of the Western Cape, Bellville, Cape Town, 7535, South Africa\label{aff161}
\and
DAMTP, Centre for Mathematical Sciences, Wilberforce Road, Cambridge CB3 0WA, UK\label{aff162}
\and
Kavli Institute for Cosmology Cambridge, Madingley Road, Cambridge, CB3 0HA, UK\label{aff163}
\and
Department of Astrophysics, University of Zurich, Winterthurerstrasse 190, 8057 Zurich, Switzerland\label{aff164}
\and
IRFU, CEA, Universit\'e Paris-Saclay 91191 Gif-sur-Yvette Cedex, France\label{aff165}
\and
Oskar Klein Centre for Cosmoparticle Physics, Department of Physics, Stockholm University, Stockholm, SE-106 91, Sweden\label{aff166}
\and
Astrophysics Group, Blackett Laboratory, Imperial College London, London SW7 2AZ, UK\label{aff167}
\and
Univ. Grenoble Alpes, CNRS, Grenoble INP, LPSC-IN2P3, 53, Avenue des Martyrs, 38000, Grenoble, France\label{aff168}
\and
INAF-Osservatorio Astrofisico di Arcetri, Largo E. Fermi 5, 50125, Firenze, Italy\label{aff169}
\and
Dipartimento di Fisica, Sapienza Universit\`a di Roma, Piazzale Aldo Moro 2, 00185 Roma, Italy\label{aff170}
\and
Centro de Astrof\'{\i}sica da Universidade do Porto, Rua das Estrelas, 4150-762 Porto, Portugal\label{aff171}
\and
HE Space for European Space Agency (ESA), Camino bajo del Castillo, s/n, Urbanizacion Villafranca del Castillo, Villanueva de la Ca\~nada, 28692 Madrid, Spain\label{aff172}
\and
Department of Astrophysical Sciences, Peyton Hall, Princeton University, Princeton, NJ 08544, USA\label{aff173}
\and
Theoretical astrophysics, Department of Physics and Astronomy, Uppsala University, Box 515, 751 20 Uppsala, Sweden\label{aff174}
\and
Minnesota Institute for Astrophysics, University of Minnesota, 116 Church St SE, Minneapolis, MN 55455, USA\label{aff175}
\and
Mathematical Institute, University of Leiden, Einsteinweg 55, 2333 CA Leiden, The Netherlands\label{aff176}
\and
School of Physics \& Astronomy, University of Southampton, Highfield Campus, Southampton SO17 1BJ, UK\label{aff177}
\and
Space physics and astronomy research unit, University of Oulu, Pentti Kaiteran katu 1, FI-90014 Oulu, Finland\label{aff178}
\and
Center for Computational Astrophysics, Flatiron Institute, 162 5th Avenue, 10010, New York, NY, USA\label{aff179}
\and
Department of Astronomy, University of Massachusetts, Amherst, MA 01003, USA\label{aff180}
\and
Department of Physics and Astronomy, University of British Columbia, Vancouver, BC V6T 1Z1, Canada\label{aff181}}